\documentclass[showpacs,aps,prb,twocolumn]{revtex4}
\usepackage{epsfig}
\usepackage{amsmath}

\def\rr{{\bf r}}
\def\xx{{\bf x}}
\def\kk{{\bf k}}
\def\pp{{\bf P}}
\def\jj{{\bf J}}
\def\aa{{\bf a}}
\def\RR{{\bf R}}
\def\bb{{\bf b}}
\def\ee{\boldsymbol{\cal E}}
\def\hh{\hat{H}}
\def\tk{\hat{T}_\kk}
\def\ket#1{\vert#1\rangle}
\def\bra#1{\langle#1\vert}

\begin{document}

\title{Dynamics of Berry-phase polarization in time-dependent electric fields}

\author{Ivo Souza}
\author{ Jorge \'I\~niguez}
\altaffiliation{Present address: NIST Center for Neutron Research and 
Department of Materials and Nuclear Engineering of the University of Maryland.}
\author{David Vanderbilt}

\affiliation{Department of Physics and Astronomy, Rutgers
University, Piscataway, New Jersey 08854-8019, USA}

\begin{abstract}

We consider the flow of polarization current $\jj=d\pp/dt$
produced by a homogeneous electric field $\ee(t)$ or by rapidly varying 
some other parameter in the Hamiltonian of a solid. 
For an initially insulating system and a collisionless time evolution,
the dynamic polarization $\pp(t)$
is given by a nonadiabatic version of the King-Smith--Vanderbilt 
geometric-phase formula.
This leads to a computationally convenient form for the Schr\"odinger equation
where the electric field is described by a linear scalar potential
handled on a discrete mesh in reciprocal space. 
Stationary solutions in sufficiently weak
static fields are local minima of the energy functional of Nunes and Gonze.
Such solutions only exist below a critical field that depends inversely
on the density of $k$ points. For higher fields they become long-lived
resonances, which can be accessed dynamically by gradually increasing $\ee$. 
As an illustration the dielectric function 
in the presence of a dc bias field is computed for a tight-binding model from
the polarization response to a step-function discontinuity in $\ee(t)$,
displaying the Franz-Keldysh effect. 

\end{abstract}
\date{\today}

\maketitle


\section{INTRODUCTION}
\label{sec:intro}

A very successful theoretical and computational framework was developed by
King-Smith and Vanderbilt\cite{ksv93} for dealing within periodic boundary 
conditions with 
the macroscopic dielectric polarization of an 
insulator. The central result of the theory of bulk polarization (TBP)
is an expression for the electronic contribution
$\pp$ which takes the form of a Berry's phase\cite{berry84} 
of the valence-band Bloch
wave functions transported across the Brillouin zone (BZ). Alternatively, it 
can be recast in real space as the vector sum of the centers of charge of the 
valence-band Wannier functions. 
Practical prescriptions were devised for computing both the Berry's 
phase\cite{ksv93} and the Wannier functions,\cite{mv97} 
which have become standard features of first-principles electronic
structure codes. 

The measurable quantity accessed by the TBP is the change $\Delta \pp$ in 
macroscopic polarization induced by changing some parameter $\lambda$ in 
the electronic Hamiltonian $\hh(t)=\hh[\lambda(t)]$. The following 
assumptions were explicitly made in the original derivation\cite{ksv93}. 
(i) Adiabaticity: the change in $\lambda(t)$ is slow enough that
the electrons remain in the instantaneous ground 
state of $\hh(t)$, apart from small deviations 
proportional to $d\lambda/dt$ described by first-order adiabatic 
perturbation theory; (ii) the ground state of $\hh(t)$ remains insulating at 
all times, separated from excited states by finite energy gaps; (iii) $\hh(t)$
is lattice-periodic.
The first two assumptions are related in that the size of the energy gap sets 
the scale for deviations from adiabaticity.

A spatially homogeneous electric field necessarily violates either
(i) or (iii): if the field is introduced via a 
vector-potential  
${\bf A}(t)=-c\int^t \boldsymbol{\cal E}(t')dt'$, $\hat{H}(t)$ remains 
lattice-periodic but changes nonadiabatically, even for a static 
field; if instead a scalar potential term
$e\boldsymbol{\cal E}(t)\cdot\hat{\bf r}$ is used, $\hat{H}(t)$ is no longer 
lattice-periodic.
Nevertheless, the TBP has been successfully applied to situations where 
electric fields are 
present,\cite{nunes94,dalcorso94,dalcorso96,nunes01,souza02,umari02,sai02}
but a
rigorous justification for doing so
is still lacking. 

In this paper we reexamine the TBP and find that it can be generalized as 
follows. Assumption (i) can be dropped altogether.
Assumption (ii)
is only 
invoked at $t=0$; the ensuing nonadiabatic dynamics
may admix considerable amounts of excited states into the occupied subspace.
Finally assumption (iii) can be relaxed
to allow for a linear scalar potential to be present in addition to the
periodic crystal potential. 

These generalizations extend the scope of the TBP to 
{\it nonadiabatic} polarization currents induced by time-dependent electric 
fields, or by other rapid changes in $\hat{H}(t)$ (e.g., the initial
nonthermal ionic motion that accompanies photoexcitation of the electrons by 
an intense laser pulse.\cite{mazur02})
The dynamical equations for the electrons 
that come out of this generalized TBP
are derived and applied in the context of a tight-binding model.
These equations are semiclassical (the electrons are treated
quantum-mechanically, whereas the electric field is treated classically)
and nonperturbative (electric fields of finite magnitude are allowed).

We begin by considering in Section~\ref{sec:dynamics_dm} some general 
properties of the 
coherent dynamics of Bloch electrons that are initially in an insulating state.
They are used in Section~\ref{sec:nonadiabatic_pol} to discuss
the macroscopic current $\jj(t)$, which is expressed as the rate of 
change of a dynamic polarization $\pp(t)$ given by nonadiabatic versions of 
the King-Smith--Vanderbilt expressions.
In Section~\ref{sec:dynamical} we derive from this generalized TBP a 
numerically convenient form for the time-dependent Schr\"odinger equation 
(TDSE) in the scalar-potential gauge, discretized on a mesh of $k$ points. 
Stable stationary solutions in static fields are discussed in 
Section~\ref{sec:stationary}. They exist only 
below a critical field ${\cal E}_c$ which decreases with
increasing $k$-point density, and are local minima
of the energy 
functional of Nunes and Gonze.\cite{nunes01,souza02,umari02} A 
prescription is given for computing them using an iterative diagonalization 
scheme.
In Section~\ref{sec:results} we show numerically on a tight-binding
model how the regime above the critical 
field can be accessed dynamically, by gradually increasing the electric field
beyond the critical value. We also compute 
the dielectric function of the same model in the presence of a static bias 
field, displaying the Franz-Keldysh effect.


\section{General properties of the dynamics}
\label{sec:dynamics_dm}

\subsection{Lattice-periodicity}
\label{sec:periodicity}

Here we expound in more detail an argument, sketched in 
Ref.~\onlinecite{souza02}, that makes use of the one-particle density matrix
to handle the presence of electric fields in a well-controlled
fashion.\cite{lazzeri03}
We say that the one-particle density matrix
$n(\rr,\rr')=\langle \rr|\hat{n}|\rr'\rangle$ is lattice-periodic if
\begin{equation}
\label{eq:dm_lattice_per}
n(\rr,\rr')=n(\rr+\RR,\rr'+\RR),
\end{equation}
where $\RR$ is a lattice vector. In particular, this implies periodicity of the
charge density. Suppose that (\ref{eq:dm_lattice_per}) is true at $t=0$ (e.g.,
the electrons are in the ground state of the crystal Hamiltonian
$\hat{H}^{0}(t=0)$).
At that time a homogeneous electric field is turned on, which 
may subsequently have an arbitrarily strong and rapid variation;
$\hat{H}^{0}(t)$ may also undergo arbitrarily rapid variations (but must
remain periodic). The full Hamiltonian in the scalar potential gauge is
\begin{equation}
\label{eq:hamiltonian}
\hat{H}(t)=\hat{H}^0(t)+\hat{H}^{\ee}(t),
\end{equation}
where $\hat{H}^{\ee}(t)=e{\ee}(t)\cdot\hat{\rr}$ 
describes the electric field in the dipole approximation,
and $-e$ is the electron charge. 

Let us show that in the absence of scattering the lattice-periodicity of 
$n(\rr,\rr')$ is preserved at all later times. 
It suffices to establish that $\dot{n}(\rr,\rr')=\dot{n}(\rr+\RR,\rr'+\RR)$.
The density matrix evolves according to
$i\hbar\, d\hat{n}/dt=[\hat{H},\hat{n}]$, or, in the position 
representation,
\begin{equation}
i\hbar\,\dot{n}(\rr,\rr')=\int [H(\rr,\xx)n(\xx,\rr')-
n(\rr,\xx)H(\xx,\rr')]d\xx.
\end{equation}
(When 
left unspecified, the domain of integration over spatial coordinates is
understood to be the entire space.)
For clarity we consider the effect of $\hat{H}^0$ and
$\hat{H}^{\ee}$ separately. The $\hat{H}^0$ term yields
\begin{eqnarray}
i\hbar\,\dot{n}(\rr+\RR,\rr'+\RR)&=&\int 
[H^0(\rr+\RR,\xx)n(\xx,\rr'+\RR)
\nonumber \\&-&n(\rr+\RR,\xx)H^0(\xx,\rr'+\RR)]d\xx.
\end{eqnarray}
Making the change of variables $\xx'=\xx-\RR$ and invoking the
lattice-periodicity of $\hat{H}^0$ and $\hat{n}$, we find
$\dot{n}(\rr+\RR,\rr'+\RR)=\dot{n}(\rr,\rr')$.
Using 
$\rr(\rr,\rr')=\langle \rr|\hat{\rr}|\rr'\rangle=\rr\,\delta(\rr-\rr')$ 
the contribution from $\hat{H}^{\ee}$  
is seen to have the same property:
\begin{eqnarray}
i\hbar\, \dot{n}(\rr+\RR,\rr'+\RR)&=&e\ee\cdot(\rr+\RR)\,n(\rr+\RR,\rr'+\RR) 
\nonumber \\&-&e\ee\cdot(\rr'+\RR)\,n(\rr+\RR,\rr'+\RR)\nonumber \\
&=&e\ee\cdot(\rr-\rr')\,n(\rr,\rr')=i\hbar\, \dot{n}(\rr,\rr').\nonumber\\
\end{eqnarray}
Hence $n(\rr,\rr')$ remains lattice-periodic under the action of the full
Hamiltonian (\ref{eq:hamiltonian}).
This was to be expected, since in the
vector potential gauge the Hamiltonian is periodic.\cite{krieger86}
The purpose of this exercise was to show explicitly
how this result comes about in the scalar-potential gauge,
where the
nonperiodicity of $\hat{H}$ has been a source of some confusion regarding
this issue.

\subsection{Wannier-representability}

The previous result on the conservation of lattice-periodicity is valid for 
both metals and insulators. In what follows we
shall specialize to the case where the system is initially in an insulating
state, in which case a stronger statement can be made regarding the nature of
the states at $t>0$.

We will assume the absence of spin degeneracy throughout, so that states are 
singly-occupied. In terms of the valence Bloch eigenstates of $\hat{H}^0(t=0)$,
the initial density matrix is
\begin{equation}
n(\rr,\rr';t=0)=\Omega_B^{-1}\sum_{n=1}^M\int d\kk\,\psi_{\kk n}(\rr)
\psi_{\kk n}^*(\rr'),
\end{equation}
where the integral is over the BZ of volume 
$\Omega_B=(2\pi)^3/v$,
and $M$ is the number of filled bands. 
(Clearly, such a density matrix is lattice-periodic. Its idempotency 
can
be checked using Eq.~(\ref{eq:normalization}).)
We shall prove that, as the density matrix evolves in time according to
\begin{equation}
\label{eq:dyn_dm}
i\hbar\,\dot{n}(\rr,\rr';t)=\bra{ \rr}[\hat{H}^0+\hat{H}^{\ee},\hat{n}]
\ket{\rr'},
\end{equation}
it can still be expressed in the same form,
\begin{equation}
\label{eq:dm_bloch}
n(\rr,\rr';t)=\Omega_B^{-1}\sum_{n=1}^M\int d\kk\,\phi_{\kk n}(\rr,t)
\phi_{\kk n}^*(\rr',t).
\end{equation}
Although at $t>0$ the occupied states $\phi_{\kk n}(\rr,t)$ may depart 
significantly from the valence states of $\hat{H}^0(t)$,
they remain orthonormal and {\it Bloch-like}:
$\phi_{\kk n}(\rr,t)=e^{i\kk\cdot\rr}v_{\kk n}(\rr,t)$, with 
$v_{\kk n}(\rr+\RR,t)=v_{\kk n}(\rr,t)$.

The cell-periodic states $v_{\kk n}(\rr,t)$ are the central
objects in our formalism. For discussion purposes only, let us 
expand them in the set of eigenstates 
$u_{\kk m}(\rr,t)=e^{-i\kk\cdot\rr}\psi_{\kk m}(\rr,t)$
of the cell-periodic Hamiltonian
$\hat{H}_\kk^0(t)=e^{-i\kk\cdot\hat{\rr}}\hat{H}^0(t)e^{i\kk\cdot\hat{\rr}}$:
\begin{equation}
\label{eq:linear_comb}
\ket{v_{\kk n}(t)}= \sum_{m=1}^{\infty} 
c_{\kk,nm}(t)\,\ket{u_{\kk m}(t)}.
\end{equation}
Individual eigenstates will in general have 
fractional occupations $0\leq n_{\kk m}=\sum_{n=1}^M |c_{\kk,nm}|^2\leq 1$
at $t>0$, but the total population 
$n_\kk=\sum_{m=1}^\infty n_{\kk m}$ is the same for every $\kk$ and equals 
the number of filled bands at $t=0$.
This is intuitively clear, since a spatially homogeneous electric field causes
vertical transitions in $k$-space which amount to a redistribution of the
electron population among states with equal $\kk$; the same is
true for the transitions induced by varying the lattice-periodic 
$\hat{H}^0(t)$.

We will justify Eq.~(\ref{eq:dm_bloch}) by deriving a dynamics for
the $|v_{\kk n}\rangle$ that insures that Eq.~(\ref{eq:dm_bloch})
provides a solution to (\ref{eq:dyn_dm}).  Since there is a gauge
freedom
\begin{equation}
\label{eq:gauge_transf}
|v_{\kk n}\rangle\rightarrow\sum_{m=1}^M\,U_{\kk,mn}|v_{\kk m}\rangle
\end{equation}
($U_\kk$ is a ${\bf k}$-dependent unitary $M\times M$ matrix)
in the definition of the $|v_{\kk n}\rangle$,\cite{mv97}
the evolution equation for them is not unique.  We require only that the
$|v_{\kk n}\rangle$ should yield the correct dynamics for the gauge-invariant 
density matrix, Eq.~(\ref{eq:dyn_dm}), and we will look for the simplest
solution that achieves this goal.

By hypothesis, at time $t$ $\dot{n}(\rr,\rr')$ takes
the form
\begin{eqnarray}
\label{eq:dm_dot}
\dot{n}(\rr,\rr')&=&\Omega_B^{-1}\sum_{n=1}^M\,\int d\kk\, 
e^{i\kk\cdot(\rr-\rr')} \,\times \nonumber\\
&&\quad
[\dot{v}_{\kk n}(\rr)v_{\kk n}^*(\rr')+v_{\kk n}(\rr)\dot{v}_{\kk n}^*(\rr')].
\end{eqnarray}
As in the previous subsection, we consider the contributions from $\hat{H}^0$
and $\hat{H}^{\ee}$ in Eq.~(\ref{eq:dyn_dm})
separately. The former is captured by 
$i\hbar|\dot{v}_{\kk n}\rangle=\hat{H}_\kk^0|v_{\kk n}\rangle$.
To deal with $\hat{H}^{\ee}$ we resort to manipulations familiar from the
crystal-momentum representation (CMR)\cite{blount62} (but with the crucial 
difference that in the CMR those manipulations are applied to the 
$\ket{u_{\kk n}}$, not to the $\ket{v_{\kk n}}$). We first observe that
\begin{eqnarray}
&&\langle \rr|[\hat{\rr},\hat{n}]|\rr'\rangle=(\rr-\rr')\,n(\rr,\rr') 
\nonumber \\
&=&\Omega_B^{-1}\sum_{n=1}^M\int d\kk\,v_{\kk n}(\rr)v_{\kk n}^*(\rr')
(-i\partial_\kk)e^{i\kk\cdot(\rr-\rr')}.
\end{eqnarray}
Integrating by parts and noting that in a {\it periodic gauge}
($\phi_{\kk+{\bf G},n}=\phi_{\kk n}$) the boundary term vanishes, we obtain
\begin{eqnarray}
&&\langle \rr|[\hat{H}^{\ee},\hat{n}]|\rr'\rangle=
\Omega_B^{-1}\sum_{n=1}^M\int d\kk\,
e^{i\kk\cdot(\rr-\rr')}ie{\ee}\cdot \nonumber \\
&&\qquad\quad\Big[
\big(\partial_\kk v_{\kk n}(\rr)\big)v_{\kk n}^*(\rr')
  +v_{\kk n}(\rr)\big(\partial_\kk v_{\kk n}^*(\rr')\big) 
\Big].
\end{eqnarray}
Comparing with Eqs.~(\ref{eq:dyn_dm}) and (\ref{eq:dm_dot}) we arrive at
$i\hbar|\dot{v}_{\kk n}\rangle=ie\ee\cdot\partial_\kk|v_{\kk n}\rangle$.
The effect of $\hat{H}^{\ee}$ thus
takes the form of a $k$-derivative, and
the combined effect of $\hat{H}^0$ and $\hat{H}^{\ee}$ is
\begin{equation}
\label{eq:tdse_continuum}
i\hbar\ket{\dot{v}_{\kk n}}=(\hat{H}^0_\kk+
ie{\ee}\cdot\partial_\kk)\ket{{v}_{\kk n}}.
\end{equation}
This is our version of the TDSE for Bloch electrons in the scalar 
potential gauge, constructed in order that 
Eq.~(\ref{eq:dm_bloch}) will satisfy Eq.~(\ref{eq:dyn_dm}). 
The time-independent version was introduced as an {\it ansatz}
in Ref.~\onlinecite{nunes01}.\cite{foot:mistake}  The equivalence  of 
Eq.~(\ref{eq:tdse_continuum}) to other forms in the literature is established
in Appendix~\ref{app:pol}.

If at time $t$ the $M$ states $|v_{\kk n}\rangle$ at every $\kk$
are lattice-periodic and orthonormal,
the dynamics dictated by Eq.~(\ref{eq:tdse_continuum}) preserves those 
properties, i.e.,
$\dot{v}_{\kk n}(\rr+\RR)=\dot{v}_{\kk n}(\rr)$ and
$d\langle v_{\kk n}|v_{\kk m} \rangle/dt=0$. This can be seen
as follows. Starting from
\begin{equation}
i\hbar\, \dot{v}_{\kk n}(\rr+\RR)=\int 
H_k^0(\rr+\RR,\xx)v_{\kk n}(\xx)\,d\xx+
ie\ee\cdot\partial_\kk v_{\kk n}(\rr+\RR),
\end{equation}
making the change of variables $\xx'=\xx-\RR$, and invoking the assumed
lattice-periodicity of both $\hat{H}^0$ and $v_{\kk n}(\rr)$, the 
right-hand-side becomes $i\hbar\dot{v}_{\kk n}(\rr)$. 
As for orthonormality, Eq.~(\ref{eq:tdse_continuum}) 
yields\cite{foot:wavepacket}
\begin{equation}
\label{eq:orthonormality}
\frac{d}{dt}\langle v_{\kk n}|v_{\kk m}\rangle=\frac{e}{\hbar}\ee\cdot
\partial_\kk\langle v_{\kk n}|v_{\kk m}\rangle.
\end{equation}
Since by hypothesis 
\begin{equation}
\label{eq:norm_conv}
\langle v_{\kk n}|v_{\kk m}\rangle\equiv \int_v v_{\kk n}^*(\rr)
v_{\kk m}(\rr)\,d\rr=\delta_{n,m}, 
\end{equation}
where the integral is over a unit cell, the
right-hand-side of Eq.~(\ref{eq:orthonormality}) vanishes. 
This completes the proof of Eq.~(\ref{eq:dm_bloch}).

Two assumptions were made in the above derivation. The first is that the states
$\ket{v_{\kk n}}$ vary smoothly with $\kk$, so that $k$-derivatives
exist; we will come back to this point in Sec.~\ref{sec:dyn_pol_disc}.
The second is that the dynamics is scattering-free.
Note that Eq.~(\ref{eq:orthonormality}) is closely related to the
collisionless Boltzmann equation; incoherent scattering would destroy the
constancy of the total population $n_\kk$ by inducing transitions between
different $k$ points.

Having established that the occupied manifold is spanned by $M$ 
Bloch-like states at each $\kk$, we now transform them
into Wannier-like states
$\langle \rr|W_{{\bf R}n}(t)\rangle=W_n(\rr-{\bf R},t)$
in the usual way,
\begin{equation}
\label{eq:wannier_like}
\ket{W_{{\rm\bf R}n}(t)}=
\Omega_B^{-1}\sum_{m=1}^M\,\int d\kk\, 
e^{-i\kk\cdot{\rm\bf R}}\,U_{\kk,mn}(t)\,\ket{\phi_{\kk m}(t)}, 
\end{equation}
where a periodic gauge is assumed and we have inserted a unitary rotation
(\ref{eq:gauge_transf}) among the occupied states.
The assumption that by a judicious choice of the matrices $U_\kk(t)$
the Bloch-like states can be made to vary smoothly with $\kk$ is equivalent
to the assumption that the Wannier-like states can be chosen to be
well localized.\cite{mv97}
 
The density matrix (\ref{eq:dm_bloch}) can now be recast as
\begin{equation}
\label{eq:w-r}
n(\rr,\rr';t)=\sum_{n=1}^M\sum_{{\rm\bf R}}
W_{{\rm\bf R}n}(\rr,t)\,W_{{\rm\bf R}n}^*(\rr',t).
\end{equation}
We will term {\it Wannier-representable} (WR) a state whose density matrix 
is of this form.  An insulating ground state is WR,
while a metallic state is not. We have established that under the Hamiltonian
(\ref{eq:hamiltonian}) and in the absence of
scattering, an initially insulating system remains WR, or 
``insulating-like'', even if at some later time the ground state of
$\hat{H}^0(t)$ becomes metallic.\cite{foot:frustrated_metal}
Unlike a true insulating ground state, or a stationary 
field-polarized state,\cite{souza02} a dynamic WR state
will in general break time-reversal symmetry and carry a macroscopic current. 
This is the subject of the next Section.


\section{Dynamic polarization and current} 
\label{sec:nonadiabatic_pol}

\subsection{Derivation}
\label{sec:dyn_pol_der}

Our aim in this Section is to show that
a WR state carries a current that can be expressed as the 
rate of change of a polarization per unit volume,
\begin{equation}
\label{eq:curr_pol}
\jj(t)=\frac{d\pp(t)}{dt},
\end{equation}
where $\pp(t)$ is given in a periodic gauge by
\begin{equation}
\label{eq:dyn_pol}
P_{\alpha}(t)=-\frac{ie}{{(2\pi)}^3}\,\sum_{n=1}^M\int d\kk\,
\langle v_{\kk n}(t)|\partial_{k_{\alpha}}v_{\kk n}(t)\rangle
\end{equation}
($\alpha$ is a cartesian direction) or, equivalently, by
\begin{equation}
\label{eq:pol_wannier}
\pp(t)=-\frac{e}{v}\,\sum_{n=1}^M\,\int\,\rr\,
\vert W_n(\rr,t)\vert^2\,d\rr.
\end{equation}
Eqs.~(\ref{eq:dyn_pol})-(\ref{eq:pol_wannier}) are identical to the 
King-Smith--Vanderbilt expressions appropriate
for the adiabatic regime and $\ee=0$,\cite{ksv93} except that in 
(\ref{eq:dyn_pol}) the 
valence-band eigenstates $\ket{u_{\kk n}}$
have been replaced by the instantaneous solutions 
of the TDSE (\ref{eq:tdse_continuum}),
and the $W_n(\rr,t)$ in Eq.~(\ref{eq:pol_wannier})
are the Wannier states corresponding to the $v_{\kk n}(\rr,t)$.
Eq.~(\ref{eq:dyn_pol}) can be interpreted as
a {\it nonadiabatic} geometric phase.\cite{aa87}

As in the adiabatic case, 
$\pp(t)$ is invariant under the transformation (\ref{eq:gauge_transf}) only 
up to a ``quantum of polarization'' $(e/v){\bf R}$.
Naturally, this gauge indeterminacy does not affect the measurable $\jj(t)$.
The total change in bulk polarization in a time interval $[0,T]$ is also 
well-defined as the integrated current: $\Delta \pp=\int_0^T \jj(t)\,dt$.
It can be determined, apart from an integer multiple of the quantum, by 
evaluating $\pp(t)$ at the endpoints: $\Delta \pp=\pp(T)-\pp(0)$.
In practice the remaining indeterminacy can be removed in the manner described
in Ref.~\onlinecite{ksv93}, by evaluating $\pp(t)$ with sufficient frequency during
that interval.

To establish Eqs.~(\ref{eq:curr_pol})-(\ref{eq:dyn_pol}), we first
evaluate $d\pp/dt$ by taking the time derivative of
Eq.~(\ref{eq:dyn_pol}) and obtain, after an integration by parts,
\begin{equation}
\label{eq:dpdt}
\frac{dP_\alpha}{dt}=-\frac{ie}{{(2\pi)}^3}\,\sum_{n=1}^M\, \int d\kk\,
\big[
  \langle \dot{v}_{\kk n}|\partial_{k_{\alpha}}v_{\kk n}\rangle-\mbox{c.c.}
\big].
\end{equation} 
Inserting the TDSE, Eq.~(\ref{eq:tdse_continuum}), we note that the
contribution arising from the second term therein, which
explicitly involves $\ee$, may be written as
\begin{equation}
\label{eq:vanishing_part}
\widetilde{J}_{\alpha}=\frac{e^2}{{(2\pi)}^3\hbar}\,\sum_{n=1}^M\sum_{\beta}\,
{\cal E}_{\beta}\int d\kk\,\Omega^{(n)}_{\alpha\beta}(\kk)
\end{equation}
where
\begin{equation}
\label{eq:curvature}
\Omega^{(n)}_{\alpha\beta}(\kk)=i\big[ 
\langle\partial_{k_{\alpha}}v_{\kk n}|\partial_{k_{\beta}}v_{\kk n}\rangle-
\langle\partial_{k_{\beta}}v_{\kk n}|\partial_{k_{\alpha}}v_{\kk n}\rangle
\big].
\end{equation}
This takes the form of a (nonadiabatic) Berry
curvature.\cite{explan-notanom} Using Stokes' theorem,
its volume integral can be turned into
a surface integral around the edges of the BZ of the Berry connection
${\bf A}_{\kk,nn}$, where
\begin{equation}
\label{eq:connection}
A_{\kk,mn}^\alpha=i\langle v_{\kk m}|\partial_{k_\alpha} v_{\kk n}\rangle.
\end{equation}
Such an integral vanishes in a periodic gauge, so that 
$\widetilde{J}_{\alpha}=0$.  The remaining contribution, arising
from the insertion of the first term of Eq.~(\ref{eq:tdse_continuum})
into Eq.~(\ref{eq:dpdt}), then gives
\begin{equation}
\label{eq:curr}
\frac{dP_{\alpha}}{dt}
  =\frac{e}{{(2\pi)}^3\hbar}\,\sum_{n=1}^M\,\int d\kk
\big[
  \langle v_{\kk n}|\hat{H}_{\kk}^0|\partial_{k_{\alpha}} v_{\kk n}\rangle
  +{\rm c.c.}
\big].
\end{equation}

On the other hand, the current is
\begin{equation}
\label{eq:current}
J_{\alpha}=-\frac{e}{v}\,{\rm Tr}_c(\hat{n}\hat{v}_{\alpha}).
\end{equation}
Here 
${\rm Tr}_c$ denotes the trace per unit cell, 
\begin{equation}
\label{eq:trace}
{\rm Tr}_c(\hat{\cal O})=\frac{1}{N}\,\int 
{\cal O}(\rr,\rr)\,d\rr,
\end{equation}
where $N$ is the (formally infinite) number of real-space cells in the system.
The velocity operator is defined as
\begin{equation}
\label{eq:velocity}
\hat{v}_{\alpha}=\frac{1}{i\hbar}\,[\hat{r}_{\alpha},\hat{H}].
\end{equation}
Inserting the Hamiltonian (\ref{eq:hamiltonian}) and using
$[\hat{r}_\alpha,\hat{H}^{\boldsymbol{\cal E}}]=0$,
\begin{equation}
\label{eq:velocity_b}
\hat{v}_{\alpha}=\frac{1}{i\hbar}\,[\hat{r}_{\alpha},\hat{H}^0].
\end{equation}
In the position representation we find, combining (\ref{eq:dm_bloch}), 
(\ref{eq:current}) and (\ref{eq:velocity_b}), 
invoking the lattice-periodicity of the integrand to replace
$(1/N)\int d\rr$ by $\int_v d\rr$,
and inserting the identity
$\hat{\bf 1}=\int d\rr'\ket{\rr'}\bra{\rr'}$,
\begin{eqnarray}
J_{\alpha}&=&
-\frac{e}{{(2\pi)}^3\hbar}\sum_{n=1}^M\,\int d\kk\,\int_v d\rr\,
\int d\rr'\,
v_{\kk n}^*(\rr')v_{\kk n}(\rr) \nonumber \\ 
&\times&
H^0(\rr',\rr)\partial_{k_{\alpha}}
e^{-i\kk\cdot (\rr'-\rr)}.
\end{eqnarray}
Integrating by parts in $k_\alpha$ 
(the boundary term vanishes in a periodic gauge),
and using
\begin{equation}
H_{\kk}^0(\rr',\rr)=
e^{-i\kk\cdot(\rr'-\rr)}H^0(\rr',\rr),
\end{equation}
$J_{\alpha}$ reduces to exactly the same expression
appearing on the right-hand side of Eq.~(\ref{eq:curr}).  This
completes the proof of Eqs.~(\ref{eq:curr_pol})-(\ref{eq:dyn_pol}) for WR
states evolving under the Hamiltonian (\ref{eq:hamiltonian}).

We note in passing that the integral on the right-hand side of
Eq.~(\ref{eq:curr}) can be recast as
\begin{equation}
\int d\kk
\big[
  \partial_{k_\alpha}
  \langle v_{\kk n}|\hat{H}_{\kk}^0| v_{\kk n}\rangle-
  \langle v_{\kk n}|\big(\partial_{k\alpha}\hat{H}_{\kk}^0\big)|v_{\kk n}
  \rangle
\big].
\end{equation}
The first term vanishes in a periodic gauge, leading to the more 
familiar-looking form
\begin{equation}
J_{\alpha}=-\frac{e}{{(2\pi)}^3}\,\sum_{n=1}^M\,\int d\kk\,
\langle v_{\kk n}|\hat{v}_\alpha(\kk)| v_{\kk n}\rangle,
\end{equation}
where 
$\hat{v}_\alpha(\kk)=(1/\hbar)
\big(\partial_{k_\alpha}\hat{H}_{\kk}^0\big)$.\cite{blount62}

The above derivations (and
indeed all the results in this paper) remain valid for nonlocal
pseudopotentials such as those used in {\it ab initio} calculations,
since the definition of the velocity as the commutator
(\ref{eq:velocity}) remains valid for such pseudopotentials.


\subsection{Discussion}
\label{sec:dyn_pol_disc}

It is remarkable that a knowledge of the wave functions at $t=0$ and
$t=T$ is sufficient to infer, to within a factor of $(e/v){\bf R}$, the net 
amount of current that flowed through the bulk in the intervening time. This 
is a direct consequence of representability by localized
Wannier functions, that is, of the 
insulating-like character of the many-electron system. 
For such systems the integral in Eq.~(\ref{eq:pol_wannier}) can be
evaluated, and it becomes possible to track the time evolution of the
electronic center of mass, i.e., of $\pp$. Indeed, the center of mass 
can be meaningfully defined within periodic boundary conditions
only for many-electron states that are localized in the manner of
insulating states.\cite{kohn64,souza00} 
Under these conditions, the history of the coherent current flow is contained 
(modulo the quantum of polarization) in
the initial and final wave functions, related by
the time evolution operator $\exp[-(i/\hbar)\int_0^T \hat{H}(t) dt]$.

This result was previously established for adiabatic 
charge flow,\cite{ksv93} under the assumption that the ground
state is separated from excited states by finite energy gaps
everywhere in the BZ. In nonadiabatic situations the occupied
manifold acquires a significant excited-state admixture, so that
it becomes impossible to identify an energy gap.
Instead, underlying the derivation in Sec.~\ref{sec:dyn_pol_der}
is a weaker assumption, namely, that the many-electron state has a
localized nature, as reflected by the ability to construct,
via Eq.~(\ref{eq:wannier_like}), Wannier functions having a finite localization
length.\cite{souza00,resta99} (Numerical calculations of the
localization length will be presented in Sec.~\ref{sec:results}.)
For instance, when taking $k$-derivatives, we assumed a  
``differentiable gauge'' for the $\ket{v_{\kk n}}$.  This is only
possible if the character of the electronic manifold changes
slowly with $\kk$, which is precisely what is measured by the
localization length.\cite{mv97} These observations are
in line with Kohn's viewpoint that the defining feature of the
insulating state is wave function localization, not the existence of
an energy gap.\cite{kohn64}


\section{Dynamical equations}
\label{sec:dynamical}

Having found the Berry-phase formula (\ref{eq:dyn_pol}) for the dynamic
polarization in the presence of a field $\ee(t)$, 
let us now use it to obtain computationally tractable dynamical equations under
the Hamiltonian (\ref{eq:hamiltonian}). The starting point is the 
observation that the dipole term $\hat{H}^{\ee}$ contributes
$-v\pp(t)\cdot\ee(t)$ to the energy per unit cell. 
An energy functional valid for periodic boundary conditions is then
obtained by expressing $\pp(t)$ via the TBP 
formulas. This program was previously carried out for insulators in static
fields,\cite{nunes94,nunes01,souza02,umari02} where 
stationary states were computed by minimizing that energy
functional after applying a regularization procedure (truncation of 
the Wannier functions in real space or discretization of $k$-space).
Our strategy for the time-dependent problem is to impose stationarity on the 
corresponding action functional. Following
Refs.~\onlinecite{nunes01,souza02}, we adopt here a $k$-space formulation,
which is particularly well-suited for numerical work.
Special emphasis will be put on
the discrete-$k$ case since this is the relevant one for numerical
implementations.

\subsection{Continuum-$k$ case}
\label{sec:td_continuum}

In the continuum-$k$ limit the TDSE may be formally obtained 
from a Lagrangian density ${\cal L}(\kk)$
such that the Lagrangian per unit cell is
$L=\Omega_B^{-1}\int d\kk\,{\cal L}(\kk)$.
For WR states under the Hamiltonian (\ref{eq:hamiltonian}) we have
\begin{equation}
\label{eq:lagrangian_density}
{\cal L}(\kk)=i\hbar \sum_{n=1}^M\,\langle v_{\kk n} | \dot{v}_{\kk n} \rangle 
-E(\kk),
\end{equation}
where
\begin{equation}
\label{eq:energy_k_density}
E(\kk)=\sum_{n=1}^M\,
\langle v_{\kk n} | \hat{H}_{\kk}^0+
ie{\ee}\cdot\partial_\kk | v_{\kk n} \rangle.
\end{equation}
Using Eq.~(\ref{eq:dyn_pol}) and defining the zero-field energy functional
\begin{equation}
\label{eq:energy0}
E^0=\Omega_B^{-1}\sum_{n=1}^M\,\int d\kk\,
\bra{v_{\kk n}}\hat{H}_\kk^0\ket{v_{\kk n}},
\end{equation}
one finds the total energy functional
\begin{equation}
\label{eq:energy_continuum}
E=\Omega_B^{-1}\int d\kk\,E(\kk)=E^{0}-
v\pp\cdot\ee.
\end{equation}
The Euler-Lagrange equation\cite{goldstein80} 
\begin{equation}
\label{eq:euler-lagrange}
\frac{d}{dt}
\frac{\delta {\cal L}}{ \langle\delta \dot{v}_{\kk n} |}+
\frac{d}{d\kk} 
\frac{\delta {\cal L}}{ \langle \delta \partial_{\kk} v_{\kk n} |}-
\frac{\delta {\cal L}}{ \langle \delta v_{\kk n} |}=0
\end{equation}
then leads to the dynamical equation (\ref{eq:tdse_continuum}).

As already mentioned, the choice of dynamical equation for the 
$\ket{v_{\kk n}}$
is not unique. An
alternative to Eq.~(\ref{eq:tdse_continuum}) is
\begin{equation}
\label{eq:tdse_continuum_cov}
i\hbar\ket{\dot{v}_{\kk n}}=\big(\hat{H}^0_\kk+
ie{\ee}\cdot\widetilde{\partial}_\kk\big)\ket{{v}_{\kk n}}.
\end{equation}
The bare derivative $\partial_\kk$ has been 
replaced by 
\begin{equation}
\label{eq:cov_der}
\widetilde{\partial}_\kk= 
\partial_\kk+i\sum_{m,n=1}^M\,{\bf A}_{\kk,mn}\ket{v_{\kk m}}\bra{v_{\kk n}},
\end{equation}
where ${\bf A}_{\kk,mn}$ is given by Eq.~(\ref{eq:connection}).
The operator $\widetilde{\partial}_\kk$ is a multiband version of the covariant
derivative\cite{fradkin} and is discussed further in Appendix~\ref{app:cov}.
Although the field-coupling term in Eq.~(\ref{eq:tdse_continuum_cov})
is no longer a scalar potential term in the strict sense, we will continue
to view it as such in a generalized sense. 

Eq.~(\ref{eq:tdse_continuum_cov}) preserves the orthonormality of the
$\ket{v_{\kk n}}$ and generates the correct dynamics for the density matrix.
(These properties rely on
$(A_{\kk}^\alpha)^{\dagger}=A_{\kk}^\alpha$, which follows from 
$\partial_{k_\alpha}\bra{v_{\kk n}}v_{\kk m}\rangle=0$.\cite{foot:wavepacket_b})
The latter is most easily seen from the dynamics of the projector
\begin{equation}
\label{eq:projector}
\hat{P}_\kk=\sum_{n=1}^M \ket{v_{\kk n}}\bra{v_{\kk n}},
\end{equation}
which completely specifies the occupied subspace at $\kk$ while being 
insensitive to unitary rotations inside that subspace.
After some algebra, it can be shown
that while the individual $\ket{v_{\kk n}}$ behave differently under
Eqs.~(\ref{eq:tdse_continuum}) and (\ref{eq:tdse_continuum_cov}),
$\hat{P}_\kk$ stays the same.

An advantage of introducing Eq.~(\ref{eq:tdse_continuum_cov}) in
place of Eq.~(\ref{eq:tdse_continuum}) is that, upon the discretization
of $k$-space, the former leads to an evolution equation at point
$\kk$ that is gauge-covariant (in the sense of transforming in the
obvious way under unitary rotations among occupied states at $\kk$ and
being invariant under such rotations at neighboring points
$\kk'$), as will become clear in the next section.


\subsection{Discrete-$k$ case}
\label{sec:td_discrete}
 
This is the relevant case for numerical work.
The Lagrangian for a uniform mesh of $N$ points in the BZ is
\begin{equation}
\label{eq:lagrangian_discrete}
L=\frac{i\hbar}{N}\sum_{n=1}^M \sum_{\kk}\,
\langle v_{\kk n}|\dot{v}_{\kk n} \rangle-E,
\end{equation}
where $E$ is the energy in an electric field,
\begin{equation}
\label{eq:energy}
E=E^0-v\ee\cdot\pp,
\end{equation}
with
\begin{equation}
\label{eq:energy0_discrete}
E^0=\frac{1}{N}\sum_{n=1}^M \sum_{\kk}\,
\langle v_{\kk n} | \hat{H}_{\kk}^0 | v_{\kk n} \rangle
\end{equation}
and a discretized expression for $\pp$ to be given
shortly.  Applying the Lagrangian equations of motion
\cite{goldstein80}
\begin{equation}
\label{eq:euler-lagrange_b}
\frac{d}{dt} 
\frac{\delta L}{ \langle \delta \dot{v}_{\kk n} |}-
\frac{\delta L}{\langle \delta v_{\kk n} |}=0
\end{equation}
yields
\begin{equation}
\label{eq:td_dva}
i\hbar\frac{d}{dt}|v_{\kk n}\rangle=\hat{H}_{\kk}^0 | v_{\kk n} \rangle
-N v \ee \cdot
\frac{\delta \pp}{\bra{\delta v_{\kk n}}}.
\end{equation}
Writing
\begin{equation}
\pp=\frac{1}{2\pi} \sum_{i=1}^3\, \aa_i\,(\pp\cdot\bb_i)
\end{equation}
where $\aa_i$ and $\bb_i$ are the direct and reciprocal lattice
vectors respectively, and defining
\begin{equation}
\label{eq:pdotb}
v\pp\cdot{\bf b}_i=-e\overline{\Gamma}_i,
\end{equation} 
the last term in Eq.~(\ref{eq:td_dva}) becomes
\begin{equation}
\label{eq:td_dvb}
+\frac{N e}{2\pi} \, \sum_{i=1}^3\, (\ee\cdot\aa_i)\,
\frac{\delta \overline{\Gamma}_i}{\langle \delta v_{\kk n} |} .
\end{equation}

According to the Berry-phase theory of polarization,\cite{ksv93}
$\overline{\Gamma}_i$ is the string-averaged 
discretized geometric phase along the $\bb_i$ direction,
\begin{equation}
\label{eq:discrete_phase}
\overline{\Gamma}_i=
-\frac{1}{N^\perp_i}\,\sum_{l=1}^{N^\perp_i}\,{\rm Im}\,
{\rm ln}\prod_{j=0}^{N_i^\parallel-1}\,{\rm det}\,S(\kk_j^{(i)},\kk_{j+1}^{(i)}).
\end{equation}
Here
$S_{mn}(\kk,\kk')=\langle v_{\kk m}|v_{\kk'n}\rangle$ is the $M\times M$ 
overlap matrix, $N^\perp_1$ is the number of strings
along ${\bf b}_1$, each containing $N_1^\parallel$ points 
$\kk_j^{(1)}=\kk_{\perp}^{(1)}+j\Delta \kk_1$,
$\kk_{\perp}^{(1)}$ is a point on the $(\bb_2,\bb_3)$ plane labeled by $l$, 
and $\Delta \kk_1=\bb_1/N_1^\parallel$.
Eqs.~(\ref{eq:pdotb}) and (\ref{eq:discrete_phase}) provide the discretized
version
of the nonadiabatic Berry-phase polarization,
Eq.~(\ref{eq:dyn_pol}). 
(A discrete-$k$ formula for the macroscopic current $\jj(t)$ is
given in Appendix~\ref{app:discr}.)
As in the continuum case, a periodic gauge is assumed.

A compact expression for 
$\delta \overline{\Gamma}_i/\bra{\delta v_{\kk n}}$
is derived in Appendix \ref{app:grad-b}
using the following notation. Let 
$\kk i\sigma=\kk+\sigma\Delta\kk_i$,
where $\sigma=\pm1$. The overlap matrix becomes $S_{\kk i\sigma,mn}=
\langle v_{\kk m}|v_{\kk i\sigma,n}\rangle$. Next we define
\begin{equation}
\label{eq:def-vtkis}
|\widetilde{v}_{\kk i\sigma,n}\rangle=\sum_{m=1}^M
{\big(S_{\kk i\sigma}^{-1}\big)}_{mn}
|v_{\kk i\sigma,m}\rangle
\end{equation} 
which is a ``dual'' of $|v_{\kk n}\rangle$ in the space of
the $|v\rangle$'s at the neighboring point $\kk i\sigma$, since
\begin{equation}
\label{eq:duality}
\langle v_{\kk n}|\widetilde{v}_{\kk i\sigma,m}\rangle=\delta_{n,m}.
\end{equation}
The $|\widetilde{v}_{\kk i\sigma,n}\rangle$ are gauge-covariant in the
sense that (i) they are invariant under
unitary rotations among the $\ket{v_{\kk i\sigma,n}}$ at any neighboring
point $\kk i\sigma$, and (ii) they transform under
unitary rotations among the $\ket{v_{\kk n}}$ in the same
manner as the $\ket{v_{\kk n}}$ themselves, i.e.,
\begin{equation}
\label{eq:gauge_cov}
\ket{\widetilde{v}_{\kk i\sigma,n}}\rightarrow
\sum_{m=1}^M\,U_{\kk,mn}\ket{\widetilde{v}_{\kk i\sigma,m}}.
\end{equation}
Then it is shown in Appendix \ref{app:grad-b} that
\begin{equation}
\label{eq:grad_Gamma}
\frac{\delta \overline{\Gamma}_i}{\bra{\delta v_{\kk n}}}=
\frac{i}{2N^\perp_i}\sum_{\sigma=\pm1}\,\sigma |\widetilde{v}_{\kk i\sigma,n}
\rangle.
\end{equation}
Combining Eqs.~(\ref{eq:td_dva}), (\ref{eq:td_dvb}), and
(\ref{eq:grad_Gamma}), and defining
\begin{equation}
\label{eq:ket_w_k}
|w_{\kk n}\rangle=\frac{ie}{4\pi}\,\sum_{i=1}^3\,N_i^\parallel
(\ee\cdot\aa_i)
\sum_{\sigma=\pm1}\sigma |\widetilde{v}_{\kk i\sigma,n}\rangle,
\end{equation}
the dynamical equation becomes
\begin{equation}
\label{eq:td_discrete_a}
i\hbar\frac{d}{dt}|v_{\kk n}\rangle=\hat{H}_{\kk}^0|v_{\kk n}\rangle
+|w_{\kk n}\rangle.
\end{equation}

Eq.~(\ref{eq:td_discrete_a}) is a discretized version of  
Eq.~(\ref{eq:tdse_continuum_cov}), i.e.,
\begin{equation}
\label{eq:cov_dis}
\ket{w_{\kk n}}\simeq ie\ee\cdot\widetilde{\partial}_\kk\ket{v_{\kk n}}.
\end{equation}
This 
is connected with the fact that
the duals provide a natural framework for writing a finite-difference 
representation of $\widetilde{\partial}_\kk\ket{v_{\kk n}}$.\cite{sai02}
In our notation,
\begin{equation}
\label{eq:cov_discrete}
\Delta \kk_i \cdot \widetilde{\partial}_\kk| v_{\kk n}\rangle\simeq
\frac{1}{2}\sum_{\sigma=\pm1}\,\sigma|\widetilde{v}_{\kk i\sigma,n}
\rangle.
\end{equation}

Both dynamical equations
(\ref{eq:tdse_continuum}) and (\ref{eq:tdse_continuum_cov}) lead to
$d\langle v_{\kk n}|v_{\kk m}\rangle/dt=0$ for WR manifolds, so that 
the time evolution of the individual states $|v_{\kk n}\rangle$
is unitary. This property is 
preserved in the discretized form (\ref{eq:td_discrete_a}), 
since
$\langle v_{\kk n}|w_{\kk m}\rangle=0$.
In order to take advantage of certain
unitary integration algorithms, it is useful to recast the term
$|w_{\kk n}\rangle$ on the right-hand-side 
as an hermitian operator acting on $|v_{\kk n}\rangle$.
For that purpose let us define
\begin{equation}
\label{eq:pkis}
\hat{P}_{\kk i\sigma}=\sum_{n=1}^M |\widetilde{v}_{\kk i\sigma,n}\rangle
\langle v_{\kk n}|,
\end{equation}
which converts an occupied state at $\kk$ into its dual at $\kk i\sigma$ and
is invariant under gauge transformations (i.e., under independent unitary
rotations among occupied states at both $\kk$ and $\kk i\sigma$).
It follows that the operator
\begin{equation}
\label{eq:wkhat}
\hat{\rm w}_{\kk}(\ee)=
\frac{ie}{4\pi}\sum_{i=1}^3\,N_i^\parallel(\ee\cdot\aa_i)
\sum_{\sigma}\,\sigma\hat{P}_{\kk i\sigma}
\end{equation}
turns $|v_{\kk n}\rangle$ into $|w_{\kk n}\rangle$, which is the property
we seek. Lastly, for the purpose of acting on $|v_{\kk n}\rangle$ the
nonhermitian $\hat{\rm w}_{\kk}$
can be replaced by 
$\hat{\rm w}_{\kk}+\hat{\rm w}_{\kk}^{\dagger}$ since 
$\hat{P}_\kk \hat{P}_{\kk i \sigma}=\hat{P}_\kk$, so that
$\hat{Q}_\kk\hat{\rm w}_{\kk}=\hat{\rm w}_{\kk}$ 
(where $\hat{Q}_\kk=1-\hat{P}_\kk$)
and therefore
$\hat{\rm w}_{\kk}^{\dagger}|v_{\kk n}\rangle=0$.
We have thus achieved our goal: Eq.~(\ref{eq:td_discrete_a}) now takes
the canonical form of a TDSE,
\begin{equation}
\label{eq:td_discrete_b}
i\hbar\frac{d}{dt}|v_{\kk n}\rangle=\tk|v_{\kk n}\rangle,
\end{equation}
with an hermitian operator on the right-hand side:
\begin{equation}
\label{eq:t_k}
\tk(\ee)=\hat{H}_{\kk}^0+\hat{\rm w}_{\kk}(\ee)+
\hat{\rm w}_{\kk}^{\dagger}(\ee).
\end{equation}
We remarked previously that in the continuum-$k$ limit 
Eq.~(\ref{eq:td_discrete_a}) reduces to Eq.~(\ref{eq:tdse_continuum_cov}).
The corresponding analysis for Eqs.~(\ref{eq:td_discrete_b})-(\ref{eq:t_k}) is
left to Appendix~\ref{app:cov}.

The operator $\hat{\rm w}_{\kk}$
appearing in Eq.~(\ref{eq:t_k}) is defined via 
Eqs.~(\ref{eq:def-vtkis}), (\ref{eq:pkis}), and (\ref{eq:wkhat}).  It
should be emphasized that it depends explicitly on the occupied
states at $\kk$ and $\kk i\sigma$.
In particular, even when $\hat{H}^0_\kk$ and $\ee$ are time-independent,
if the occupied manifolds at $\kk$ and $\kk i\sigma$ are changing over time, so
is $\tk$. However, $\tk$ remains invariant under unitary rotations
at $k$-points $\kk$ and $\kk i\sigma$.
Hence the resulting dynamics of
the occupied manifold (Eq.~(\ref{eq:td_projector}) below) has the essential
property of being insensitive to the gauge arbitrariness that is always present 
in numerical simulations.

Note that
when the Lagrangian procedure was applied in Sec.~\ref{sec:td_continuum} to
the continuum-$k$ problem,
we arrived at Eq.~(\ref{eq:tdse_continuum}), which contains 
$\partial_\kk\ket{v_{\kk n}}$. When the same was done after discretization, the
resulting dynamical equation contained instead 
$\widetilde{\partial}_\kk\ket{v_{\kk n}}$. The 
reason is that the gradient of the discretized Berry's phase, 
Eq.~(\ref{eq:grad_Gamma}), is by construction orthogonal to the 
occupied subspace at $\kk$,
whereas the corresponding continuum-$k$ term used in 
Sec.~\ref{sec:td_continuum}
was not. Had we orthogonalized that term, we
would have obtained $\hat{Q}_\kk\partial_\kk\ket{v_{\kk n}}$ instead of
$\partial_\kk\ket{v_{\kk n}}$, which is
equivalent to $\widetilde{\partial}_\kk\ket{v_{\kk n}}$
(see Appendix~\ref{app:cov}).


\subsection{Numerical time integration}
\label{sec:integration}

In the applications of Sec.~\ref{sec:results} we use the
algorithm\cite{koonin90}

\begin{equation}
\label{eq:time_evolution}
\ket{v_{\kk n}(t+\Delta t)}= \frac{1-i\hbar(\Delta t/2)\hat{T}_{\kk}(t)}
{1+i\hbar(\Delta t/2)\hat{T}_{\kk}(t)}\,
\ket{v_{\kk n}(t)}
\end{equation}
to perform the time integration.
Note that in order to use this algorithm it was necessary to invoke
the form (\ref{eq:td_discrete_b}) of the TDSE.
The hermiticity of $\hat{T}_{\kk}$ guarantees that the time evolution is
strictly unitary for any value of $\Delta t$.
Since the system under study in Sec.~\ref{sec:results} is a tight-binding model
with only three basis orbitals, the matrix inversion
is very inexpensive. 
The same algorithm has been successfully used to perform
self-consistent time-dependent density-functional calculations of the
optical properties of atomic clusters using localized  
orbitals as a basis set.\cite{tsolakidis02}
For calculations with large basis sets
(e.g., plane-waves) more efficient algorithms are 
available.\cite{sugino99,watanabe02}

Owing to the hermiticity of $\tk$, the projector
(\ref{eq:projector}) obeys
\begin{equation}
\label{eq:td_projector}
\frac{d\hat{P}_\kk}{dt}=\frac{1}{i\hbar}\,[\tk,\hat{P}_\kk].
\end{equation}
Hence, a variation of the above approach would be to replace
$\tk$ by 
\begin{equation}
\hat{\cal T}_\kk=\hat{Q}_\kk \tk + \tk \hat{Q}_\kk,
\end{equation}
which is also hermitian. Because
$[\hat{\cal T}_\kk,\hat{P}_\kk]=[\tk,\hat{P}_\kk]$,
this choice does not change the dynamics of the occupied subspace
$\hat{P}_\kk$, but it does change the dynamics of the individual states
$|v_{\kk n}\rangle$.  In fact, $\hat{\cal T}_\kk$ generates
a {\it parallel transport} evolution characterized by
$\langle v_{\kk n}|\dot{v}_{\kk n}\rangle=0$, thus discarding the
``irrelevant'' part of the dynamics associated with phase factors and
unitary rotations inside the occupied subspace.
We have found empirically, however, that the use of $\hat{\cal T}_\kk$ in 
Eq.~(\ref{eq:time_evolution}) appears to result in a less stable
numerical time evolution, and we have therefore chosen to retain the
original $\tk$ dynamics in our practical implementation.


\subsection{Discussion}
\label{sec:dynamics_discussion}

It may seem surprising that a linear potential can be accommodated in a
theoretical description of a periodic bulk system.
A commonly-held viewpoint is that  a linear potential can be
implemented within periodic boundary conditions only for the case
of a finite system (molecule or cluster) in a supercell, in which
case it becomes possible to introduce a sawtooth potential as long as its
discontinuity is located in a region of negligible electron density.
To the contrary, Eqs.~(\ref{eq:tdse_continuum}), 
(\ref{eq:tdse_continuum_cov}), and (\ref{eq:td_discrete_b}) 
demonstrate that it is perfectly permissible to insist on the usual periodic
boundary conditions on the wave functions while allowing for
nonperiodicity of the potential. This can be done because
the potential takes the special form of a sum of spatially periodic and
linear contributions, relevant to a crystal in a homogeneous
electric field.  As shown in Sec.~\ref{sec:periodicity}, the 
action of the nonperiodic Hamiltonian, Eq.~(\ref{eq:hamiltonian}), then
preserves the lattice periodicity of the density matrix,
which can therefore be represented by periodic wave functions.

Incidentally, we note that 
a sawtooth operator of sorts is ``hiding'' behind the TBP
formulas. The Berry-phase polarization has been recast 
as the expectation value of a properly-defined center-of-mass 
position operator of the many-electron periodic system.\cite{souza00} That 
operator, 
introduced by Kohn,\cite{kohn64} is a sawtooth, not in real space but
in the configuration space of the many-body wave function. 
It can only be constructed for wave functions having a certain
disconnectedness (localization) property in configuration space
characteristic of the insulating state.
This observation is closely related to the discussion in
Sec.~\ref{sec:dyn_pol_disc}.

Finally, we mention that an alternative approach for introducing a linear 
potential into a periodic solid
is via the crystal momentum representation (CMR) formalism.\cite{blount62}  
This approach is summarized in
Appendix~A, where the connection with our formalism is established.
The CMR dynamical equations appear to be
less convenient for computational work.
However, the advantages of the present formulation
came at the expense of generality,
since our equations are restricted to the scattering-free dynamics of
initially insulating systems.


\section{Stable Stationary solutions}
\label{sec:stationary}

\subsection{Formulation}
\label{sec:stat_formulation}

Let us try to find, for a constant $\ee\not=0$, 
solutions to Eq.~(\ref{eq:td_discrete_b}) 
for which the occupied manifold remains unchanged over time. A natural 
guess is the manifold spanned by $M$ eigenstates of $\tk$ at each 
$\kk$,
\begin{equation}
\label{eq:eigenequation_tk}
\tk|v_{\kk n}\rangle=E_{\kk n}(\ee)|v_{\kk n}\rangle.
\end{equation}
Since $\tk$ depends on the occupied states at the neighboring $k$-points, 
Eq.~(\ref{eq:eigenequation_tk})  must be solved 
self-consistently among all $\kk$. If a solution exists,
the corresponding
$\hat{T}_{\kk}$ and $\hat{P}_{\kk}$ commute and, according to
Eq.~(\ref{eq:td_projector}), $d\hat{P}_{\kk}/dt=0$, i.e., the solution is stationary.

We are now ready to make contact with Refs.~\onlinecite{souza02,umari02}, where
the energy functional $E$ of Nunes and Gonze,\cite{nunes01} 
Eq.~(\ref{eq:energy}), was minimized at fixed
$\ee$. A stationary point of $E$ has zero gradient: 
$|G_{\kk n}\rangle=\delta E/\langle \delta v_{\kk n}|=0$,
where the functional derivative is taken in such a way that the gradient
is orthogonal to the occupied space.
In Appendix~\ref{app:grad-b} it is shown that
\begin{equation}
\label{eq:gradient_energy}
|G_{\kk n}\rangle=
(1/N)\,\hat{Q}_\kk\tk\,|v_{\kk n}\rangle,
\end{equation}
so that solutions of Eq.~(\ref{eq:eigenequation_tk}) obey
$|G_{\kk n}\rangle=0$.  Thus, stationary solutions
of the dynamical equation are
stationary points of $E$. 

A Hessian stability analysis\cite{unpublished} 
shows that a necessary condition for a stationary
point of $E$ to be a minimum is that the $M$ lowest-lying eigenstates of 
$\tk$ are chosen.
Since doing so at $\ee=0$  yields the 
ground state, at finite $\ee$ that procedure yields a state that is
is adiabatically connected to it by slowly ramping up
the field, keeping the system in a minimum of $E$.
Such ``polarized manifolds'' have been 
discussed previously in a perturbative 
framework,\cite{nenciu91,wannier55,adams57}
treating $\kk$ as a continuous variable. In that limit the electric 
field perturbation becomes singular.
That is, even an arbitrary small field induces a 
current via Zener tunneling to higher bands, and the polarized manifolds
are not stationary, but rather, are long-lived resonances.
In other words, an infinite crystal in the presence of a static electric field
does not have a ground state. This is reflected in $E$ loosing its 
minima as soon as $\ee$ departs from zero. 

Instead, for a discrete mesh of $k$-points, arguments can be 
given\cite{nunes01,souza02} suggesting that $E$ loses its minima only when 
$\ee$ exceeds a critical value ${\cal E}_c(N)$ that decreases as 
the number $N$ of $k$ points increases; this is supported by numerical 
calculations.\cite{souza02,umari02} It follows from the preceding discussion 
that the minima of $E$ below ${\cal E}_c(N)$ are stable
stationary solutions of the dynamical equation.
Conversely, above ${\cal E}_c(N)$ there are no 
such solutions. 

These two regimes --~below and above the
critical field~-- will be explored numerically in 
Sec.~\ref{sec:efield_turn_on} via time-dependent calculations. If one stays
below ${\cal E}_c(N)$, the stationary solutions can be computed
using time-independent methods, such as the diagonalization algorithm
described next or the minimization methods of 
Refs.~\onlinecite{souza02,umari02}.


\subsection{Diagonalization algorithm}
\label{sec:algebraic}

We have in Eq.~(\ref{eq:eigenequation_tk}) the basis for an algebraic method 
of computing stationary states at finite $\ee$ on a uniform
$k$-point mesh, for $|\ee|<{\cal E}_c(N)$: loop over the $k$ points; for 
each one select the $M$ eigenstates of $\tk$ with the
lowest eigenvalues; iterate until the procedure
converges at all $\kk$ and the occupied subspace stabilizes (this will only
happen below ${\cal E}_c$). 
Even in a tight-binding model without charge self-consistency, the set of
equations (\ref{eq:eigenequation_tk}) has to be solved self-consistently
throughout the BZ,
since the operators $\tk$ couple neighboring $k$ points via their
dependence on the $\ket{v_{\kk n}}$. One may choose to update 
$\tk$ either inside or outside the loop over $\kk$; the latter
option renders the algorithm parallelizable over $k$ points.

We have tested this scheme on the tight-binding model of 
Sec.~\ref{sec:results}, and confirm that it produces the same state 
as a direct steepest-descent or conjugate-gradients minimization of 
the functional $E$.\cite{souza02}
This algorithm may be especially suited for implementation in certain 
total-energy codes
that are based on iteratively diagonalizing the
Kohn-Sham Hamiltonian expanded in a small basis set of local 
orbitals.\cite{soler02}

\subsection{Discussion}
\label{sec:stat_discussion}

Eq.~(\ref{eq:eigenequation_tk}) is a discretization of the time-independent
version of Eq.~(\ref{eq:tdse_continuum_cov}):
\begin{equation}
\label{eq:stationary_continuum_cov}
\big(\hat{H}^0_\kk+
ie{\ee}\cdot\widetilde{\partial}_\kk\big)\ket{{v}_{\kk n}}=
E_{\kk n}(\ee)\ket{{v}_{\kk n}}.
\end{equation}
An analysis of the eigenvalues of this equation will serve as a guide for 
discussing those of Eq.~(\ref{eq:eigenequation_tk}).
(For the present purposes we will assume that the continuum form 
(\ref{eq:stationary_continuum_cov})
has solutions for $\ee\not=0$.) As a result of the properties of the
covariant derivative (Appendix~\ref{app:cov}) the $E_{\kk n}(\ee)$ are 
invariant under diagonal gauge transformations
$U_{\kk,mn}=e^{i\theta_{\kk m}}\delta_{m,n}$. 
Upon multiplying on the left by
$\bra{v_{\kk n}}$ the second term on the left-hand-side of
Eq.~(\ref{eq:stationary_continuum_cov}) vanishes. Integrating 
over $\kk$ and summing over $n$, we then find
\begin{equation}
\label{eq:inequality}
\Omega_B^{-1}\sum_{n=1}^M\,\int d\kk\, E_{\kk n}(\ee)=
E^0(\ee)\geq E^0(\ee=0),
\end{equation}
where $E^0(\ee)$ is the zero-field energy functional
(\ref{eq:energy0}) evaluated at the field-polarized stationary state, and the 
inequality follows from the variational principle. 
The same properties hold 
for the eigenvalues of the discretized form~(\ref{eq:eigenequation_tk}),
which can be obtained by diagonalizing $\hat{H}_\kk^0$ inside the occupied
manifold.
We have here the interesting situation that a minimum of the
{\it total} energy $E$ can be obtained by solving the eigenvalue equations
(\ref{eq:eigenequation_tk}) whose eigenvalues, summed over $n$ and $\kk$
give instead the {\it zero-field} contribution $E^0$. This can be traced 
back to Eq.~(\ref{eq:orthogonal}), which expresses the
``parallel-transport-like'' nature of the covariant derivative.

The above is to be compared with the time-independent
version of Eq.~(\ref{eq:tdse_continuum}),
\begin{equation}
\label{eq:stationary_continuum}
\big(\hat{H}^0_\kk+
ie{\ee}\cdot\partial_\kk\big)\ket{{v}'_{\kk n}}=
E'_{\kk n}(\ee)\ket{{v}'_{\kk n}}.
\end{equation}
Under a diagonal transformation
$\ket{{v}'_{\kk n}}\rightarrow e^{i\theta_{\kk m}}\,\ket{{v}'_{\kk n}}$
its eigenvalues change as
$E'_{\kk n}(\ee) \rightarrow E'_{\kk n}(\ee)-e\ee\cdot\partial_\kk
\theta_{\kk n}$. The analog of Eq.~(\ref{eq:inequality})
is
\begin{equation}
\Omega_B^{-1}\sum_{n=1}^M\,\int d\kk\, E'_{\kk n}(\ee)=
E^0(\ee)-v\ee\cdot\pp(\ee).
\end{equation}
The quantity on the right-hand side is now the total energy $E$,
which is invariant only modulo $e\ee\cdot{\bf R}$.

Although the individual eigenstates of Eq.~(\ref{eq:stationary_continuum_cov})
are in general different from those of Eq.~(\ref{eq:stationary_continuum}),
the self-consistent solutions for all $\kk$ and $n$ span the same
space in both cases, i.e., they differ only by a gauge transformation.
It is then a matter of convenience to choose which of the two
equations to solve in practice. 
Our particular approach is to discretize
Eq.~(\ref{eq:stationary_continuum_cov}) 
in a gauge-covariant manner, and then solve the resulting 
Eq.~(\ref{eq:eigenequation_tk}).


\section{Numerical results}
\label{sec:results}


\subsection{Tight-binding model}
\label{sec:tb}

\begin{figure}
\centerline{\epsfig{file=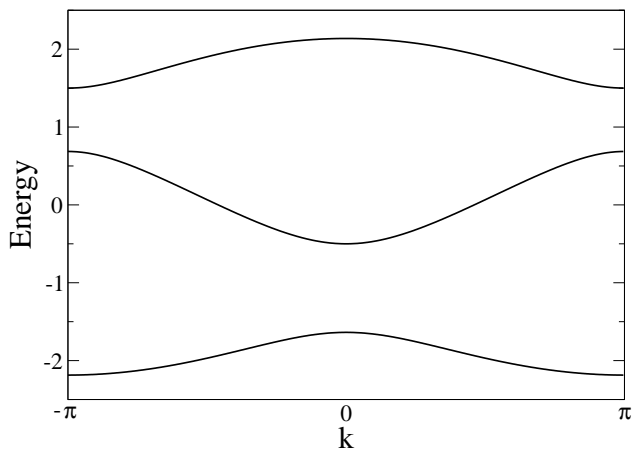,width=3.3in,angle=0}}
\caption{Energy dispersion of the tight-binding model for the choice of
parameters $t=1$, $\Delta=-1$, and $\alpha=0$.}
\label{fig1}
\end{figure}

We have applied our scheme to the one-dimensional tight-binding model 
of Ref.~\onlinecite{nunes94}, a three-band Hamiltonian with three 
atoms per unit cell of length $a=1$ and one orbital per atom,
\begin{equation}
\label{eq:tb_ham}
\hat{H}^0(\alpha)=\sum_j\,
\big\{
  \epsilon_j(\alpha)\hat{c}_j^\dagger \hat{c}_j+
  t\big[\hat{c}_j^\dagger \hat{c}_{j+1}+\rm{h.c.}\big]
\big\},
\end{equation}
with the site energy given by $\epsilon_{3m+l}(\alpha)=
\Delta\cos(\alpha-\beta_l)$. Here $m$ is the cell index, 
$l=\{ -1,0,1\}$ is the site index, and $\beta_l=2\pi l/3$.
Before the
Berry-phase polarization can be computed\cite{bennetto96} (or an electric field
applied to the system\cite{nunes94,dunlap86}), the position operator must
be specified.
Although this may be done without introducing additional 
parameters,\cite{foreman02} we adopt the simple prescription
of Ref.~\onlinecite{nunes94}:
$\hat{x}=\sum_j\, x_j\hat{c}_j^\dagger \hat{c}_j$, with $x_j=j/3$.
In the results reported below we have set $e=\hbar=1$, $t=1$, and $\Delta=-1$,
and only the lowest band is filled (with single occupancy). 
Fig.~\ref{fig1} shows the band structure at zero field for $\alpha=0$.


\subsection{Sliding charge-density wave}
\label{sec:cdw}

The Hamiltonian of Eq.~(\ref{eq:tb_ham}) is a simple model of a
commensurate charge-density wave which slides by one period as the parameter
$\alpha$ evolves adiabatically through $2\pi$.
It is easiest to see this by noting that in the space of parameters
$\Delta_x=\Delta\cos\alpha$ and $\Delta_y=\Delta\sin\alpha$, cycling
$\alpha$ by $2\pi$ corresponds to tracing a circle
about the origin in the $\Delta_x$--$\Delta_y$ plane.  The system
is insulating (i.e., a gap remains open) at all points in this plane
except for a singular point at the origin where the system is metallic.
Thus, this cyclic adiabatic change in $\hat{H}^0$ takes the 
system along an insulating path that encircles this singular point,
so that a quantized particle transport $\Delta P=\int_0^T
J(t)\,dt$ of a unit charge is obtained.\cite{thouless83}

Away from the adiabatic regime, deviations from exact quantization are 
expected.
This can be understood from the fact that under nonadiabatic conditions 
the state at time $t$ depends on the history at times $t'<t$.
In particular, the final state may be be 
different from the initial one even though $\hat{H}^0(T)=\hat{H}^0(0)$, in
which case $P(T)-P(0)\not=1$.
By contrast, an adiabatically-evolving system has no memory, being
completely determined by the instantaneous $\hat{H}^0(t)$ and $d\hat{H}^0/dt$.

To illustrate this point, we increased $\alpha$ from $0$ to $2\pi$ 
during a time interval  $t\in[0,T]$ according to
$\alpha(t)=2\pi \sin^2(\pi t/2T)$, and held it
constant afterwards. The system was prepared at $t=0$ in its ground state, and 
the wave functions evolved in time 
according to Eq.~(\ref{eq:time_evolution}), using $\Delta t=0.005$ (the same
time step was used in all other simulations in this work). 
At each time step we computed the dynamic 
polarization using Eq.~(\ref{eq:discrete_phase}).
\begin{figure}
\centerline{\epsfig{file=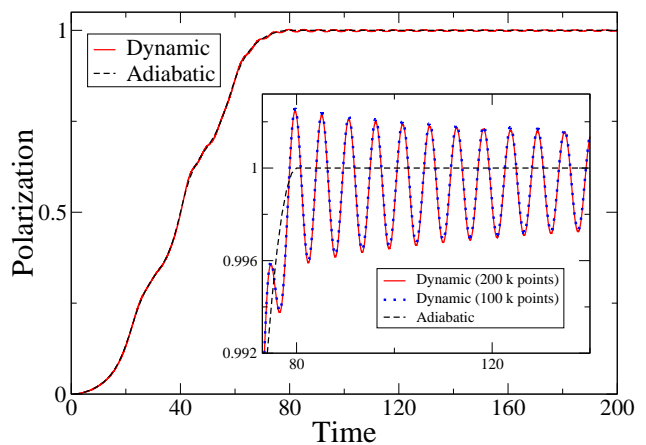,width=3.3in,angle=0}}
\vspace{0.2cm}
\caption{Time evolution of the polarization as a result of changing the sliding
parameter $\alpha$ from $0$ to $2\pi$ over the time interval $[0,80]$, using
200 $k$ points. The 
solid line shows the actual dynamic polarization, while the dashed line shows 
the ground-state polarization of the instantaneous Hamiltonian.
Inset: Detail of the remnant oscillations of the polarization after the
Hamiltonian stops changing, at $t=80$. For comparison, the results using
100 $k$ points are also shown.}
\label{fig2}
\end{figure}

\begin{figure}
\centerline{\epsfig{file=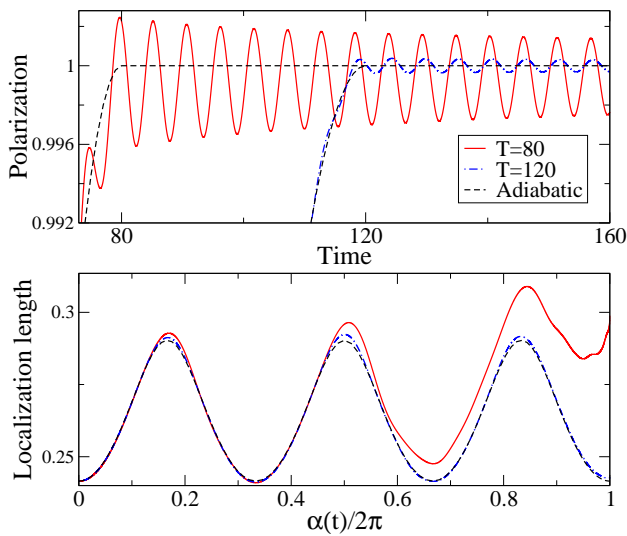,width=3.3in,angle=0}}
\vspace{0.2cm}
\caption{Upper panel: Same as Fig.~\ref{fig2}, but now using two different
time intervals, $[0,80]$ and $[0,120]$, for changing the sliding parameter
$\alpha$.
 Lower panel: electron localization 
length $\xi(t)$ versus the instantaneous value of $\alpha$, during the
time intervals $[0,T]$ during which $\alpha$ is changing.}
\label{fig3}
\end{figure}

The resulting $P(t)$ for $T=80$ and 200 $k$ points is shown in Fig.~\ref{fig2},
where we also display the exact adiabatic ($T\rightarrow\infty$) limit
$P_{\rm static}[\alpha(t)]$ obtained by diagonalizing 
$\hat{H}^0_k(\alpha(t))$ on the same mesh of $k$ points.\cite{foot:thouless}
(To check that our calculations are converged with respect to the number of 
$k$ points, the inset of
Fig.~\ref{fig2} compares the results for 100 and 200 $k$ points.)
The dynamic polarization $P(t)$ obtained by solving the TDSE follows closely, 
but not exactly, the adiabatic curve. In particular, at the time $t=80$ when 
the Hamiltonian stops changing, the polarization differs 
slightly from unity (see inset in Fig.~\ref{fig2}), indicating that the system is
not in the ground state. The oscillations that
follow arise from quantum interference
(beats) between valence and conduction states, as a 
result of having excited electrons across the gap during $[0,T]$. Their period,
of 5.5 time units, corresponds to the fundamental gap in 
Fig.~\ref{fig1}, $E_{\rm gap}=1.137$. This is consistent with
the $k$-space distribution of the (small) electron-hole pair
amplitude present in the system after time $T$: for a sliding period of
$T=80$, the distribution 
is mostly 
concentrated around $k=0$, and it is essentially the lowest conduction band 
that gets populated. 
As the adiabatic limit is approached by increasing $T$, the amplitude of the
remnant oscillations of the polarization decreases. This is illustrated in
the upper panel of Fig.~\ref{fig3}, where we compare $T=80$ with $T=120$.

Besides the macroscopic polarization $P(t)$, another quantity of interest
is the electronic localization length $\xi(t)$ that 
characterizes the root-mean-square quantum fluctuations of the macroscopic
polarization.\cite{souza00} It is given
by $\xi^2=\Omega_{\rm I}/M$, where $\Omega_{\rm I}$ is a
gauge-invariant quantity which in one dimension is equal to the
spread of the maximally-localized Wannier functions
(\ref{eq:wannier_like}).\cite{mv97} We have computed $\Omega_{\rm I}$ using 
Eq.~(34) from Ref.~\onlinecite{mv97}, and in the lower panel of Fig.~\ref{fig3}
we plot $\xi(t)$ against $\alpha(t)/2\pi$.
In the adiabatic limit the resulting curve consists of three identical 
oscillations, reflecting the existence of three equivalent atoms in the unit 
cell. As nonadiabaticity 
increases, $\xi$ tends to increase as well. Nevertheless, the electrons remain 
localized, i.e., ``insulating-like'', in the sense discussed in
Sec.~\ref{sec:dyn_pol_disc}.\cite{foot:insulating-like}

The above results are representative of the regime where deviations
from adiabaticity are small. If we increase the degree of
nonadiabaticity by choosing a smaller $T$ (e.g., $T=40$), we begin
to notice a linear increase of the polarization at later
times.  This new behavior can be traced to the excitation of
electron and hole wavepackets centered at some $k_0$ and
propagating at different group velocities.  Let $\Delta E_{k_0}$ be
the interband separation, and $\Delta v_g$ be the difference of
group velocities of the two lowest bands, at $k_0$.  Then, in addition to
the quantum beats of period $2\pi\hbar/\Delta E_{k_0}$ caused by
the interband dynamics, we observe a linear-in-$t$ term in $P(t)$
with slope proportional to $\Delta v_g$, reflecting the change in
dipole moment as the electron-hole pair separates.  (More
precisely, the preceeding statements apply only in the limit of a
dense $k$-point mesh; for any finite mesh spacing $\Delta k$, the
linear behavior is replaced by an oscillatory one with an amplitude
scaling as $1/\Delta k$ and period $2\pi/(\Delta v_g\,\Delta k)$. 
Thus, an especially fine $k$-point mesh should be used 
if these effects are to be investigated.)


\subsection{Gradual turn-on of an electric field}
\label{sec:efield_turn_on}

In the previous example the electric field was held at zero,
and the dynamics was
produced by varying the parameter $\alpha$ in $\hat{H}^0$. Let us now study the
polarization response of the system when an electric field ${\cal E}(t)$  is 
switched on linearly over a time interval $[0,T]$ and is held fixed afterwards.
We have set $\alpha=0$, so that the ground state is centrosymmetric, with zero 
spontaneous polarization.

\begin{figure}
\centerline{\epsfig{file=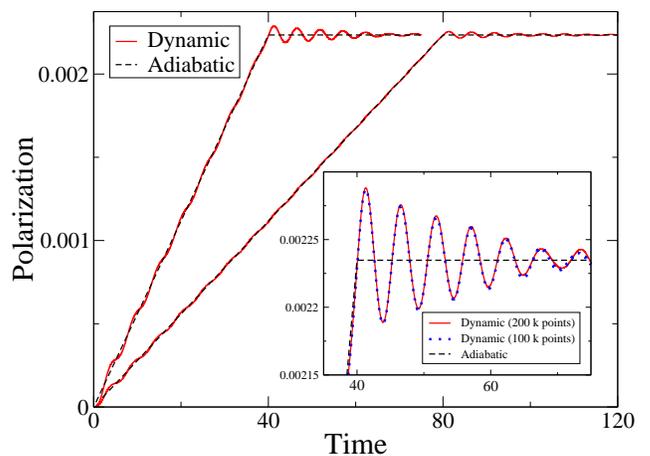,width=3.3in,angle=0}}
\caption{Time evolution of the polarization as a result of increasing the 
electric field from $0$ to ${\cal E}_{\rm max}=0.025$ over two time intervals,
$[0,40]$ and $[0,80]$, using 200 $k$ points.
The solid line shows the actual dynamic polarization, while the dashed line 
shows the static polarization for the instantaneous value of the field.
Inset: Comparison of the dynamic polarization for 100 and 200 $k$ points.}
\label{fig4}
\end{figure}

We begin by considering a situation where the final 
value of the field, ${\cal E}_{\rm max}$, is smaller than the 
$k$-mesh-dependent critical field ${\cal E}_c(N)$
above which the energy functional (\ref{eq:energy}) has no
minima. This allows us to compare the dynamic polarization $P(t)$ with the
static polarization $P_{\rm static}[{\cal E}(t)]$ of the stationary state in 
the presence of the same field, which we find by minimizing the 
energy.\cite{souza02} In Fig.~\ref{fig4} we display the results for 
${\cal E}_{\rm max}=0.025$ and two
different switching times, $T=40$ and $T=80$. The simulation was done using 200
$k$ points, to which corresponds a critical field 
${\cal E}_c(N=200)\sim 0.037$.
(The inset shows the agreement between the results obtained
using 100 and 200 $k$ points.)
Clearly, $P(t)$ tracks quite closely the adiabatic curve 
$P_{\rm static}[{\cal E}(t)]$, the more so as $T$ increases. This 
illustrates the point, 
emphasized in Ref.~\onlinecite{nunes94}, that the
state obtained by minimizing a field-dependent energy functional should be 
thought of as
the one which is generated from the zero-field state by adiabatically turning
on ${\cal E}$. 

Let us now explore the regime above ${\cal E}_c(N)$, where energy-minimization 
schemes fail. For ${\cal E}_{\rm max}>{\cal E}_c(N)$ the exact adiabatic limit
of the process of ramping up the field is unattainable. Nevertheless, if 
${\cal E}_{\rm max}$ is small compared to the field scale at which
{\it intrinsic} breakdown occurs (i.e., at which the Zener tunneling
rate becomes on the of order interband frequencies, which is a bulk property
\cite{odwyer73}), a {\it quasistationary} state should be reachable
by turning on the field at a rate that is slow compared to the
usual electronic processes, but fast compared to the characteristic
tunneling rate at the maximum field encountered.  After the ramp-up
is completed, but at times still short compared to the tunneling
rate, this state should provide the appropriate extrapolation to
fields above ${\cal E}_c(N)$ of the truly stationary state that
exists below ${\cal E}_c(N)$.

\begin{figure}
\centerline{\epsfig{file=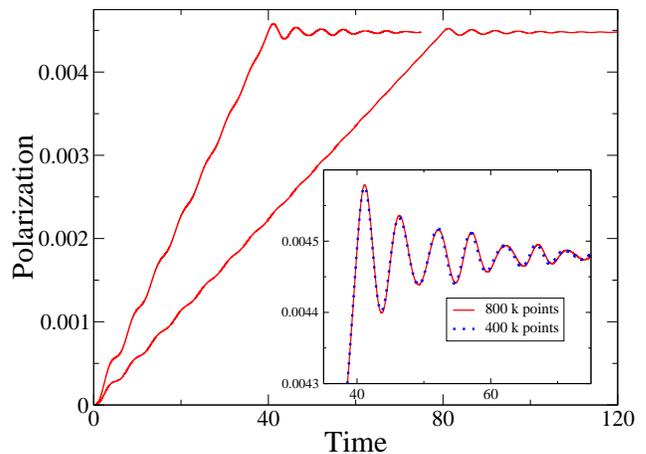,width=3.3in,angle=0}}
\caption{Same as Fig.~\ref{fig4}, but now using
${\cal E}_{\rm max}=0.05$ and 800 $k$ points. Since ${\cal E}_{\rm max}$ is
larger than the critical field for this number of $k$ points
(${\cal E}_c\sim 0.01$), no adiabatic curve $P_{\rm static}[{\cal E}(t)]$ is
shown. Inset: Comparison of the dynamic polarization for 400 and 800 $k$ 
points.}
\label{fig5}
\end{figure}

To illustrate this situation, we repeated the calculation with 200 
$k$ points depicted in Fig.~\ref{fig4}, but increasing the maximum field from 
0.025 to 0.05, somewhat larger than ${\cal E}_c(200)\sim 0.037$. The resulting 
curve for $P(t)$ is very similar
to that in Fig.~\ref{fig4}, without any sign of runaway behavior.
As a more striking example, we show in Fig.~\ref{fig5} the outcome of
calculations with the same final field of ${\cal E}_{\rm max}=0.05$, but with 
even denser sets of 400 and 800 $k$ points. For the latter ${\cal E}_c\sim0.01$,
considerably smaller than ${\cal E}_{\rm max}$, and still there is no sign of
instability. (Note also that the $P(t)$ curve in Fig.~\ref{fig5} --~whose 
vertical scale differs from that in Fig.~\ref{fig4}  by
the same factor of two that exists between the respective values of 
${\cal E}_{\rm max}$ --~looks almost identical to that in Fig.~\ref{fig4}.)
These results confirm that, as long as we are solving
a {\it time-dependent} Schr\"odinger equation for a given history of
switching on the field, there is no such thing as a $\Delta k$-dependent
critical field; the thermodynamic limit of an infinitely dense $k$-point mesh
is perfectly well defined. The only breakdown behavior that may  
be observed in short time scales
is the physical one that occurs when the applied field is
large enough that the Zener tunneling rate becomes 
significant.\cite{kane59,odwyer73} 
The concept of a $\Delta k$-dependent critical field applies only to the
attemp to obtain solutions in the presence of a static electric field from
an energy variational principle. By going back to the original dynamical 
problem of slowly ramping up the field, we circumvent the difficulties that 
ultimately resulted from trying to treat as a (stable) stationary state what is
really a long-lived resonance. 


\subsection{Dielectric function in a static field}
\label{sec:franz_keldysh}

There is great interest in modulating the optical properties of crystals and
superlattices by applying static electric fields. An example
of such an electro-optical effect is the modification of the dielectric
function. This is known as the Franz-Keldysh effect, or electroabsorption.
Although it has been extensively studied
in bulk semiconductors,\cite{yu96} quantum wells,\cite{rink89} and 
superlattices,\cite{ando97} we are not aware of any first-principles 
investigations. The present method may provide a route to such calculations.
\begin{figure}
\centerline{\epsfig{file=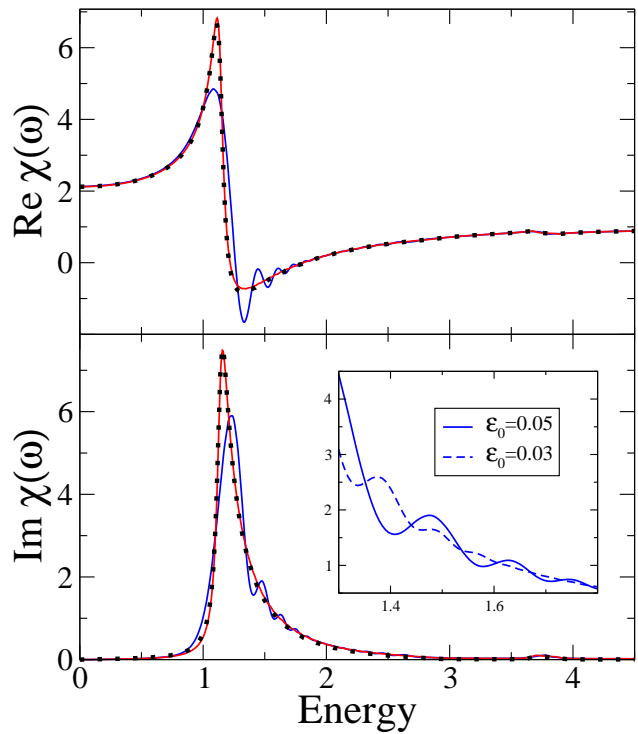,width=3.3in,angle=0}}
\caption{Susceptibility $\chi^{[{\cal E}_0]}(\omega)$ in the presence of a static
field ${\cal E}_0$, for $\alpha=0$ and 100 
$k$ points, using a level-broadening $\delta=0.04$. 
Dotted lines: Kubo formula
result for ${\cal E}_0=0$; solid lines: 
results using our method, for both ${\cal E}_0=0$ and ${\cal E}_0=0.05$.
The latter displays the Franz-Keldysh effect. The inset compares 
the Franz-Keldysh oscillations for two different bias fields, ${\cal E}_0=0.05$
and ${\cal E}_0=0.03$.}
\label{fig6}
\end{figure}

We compute the dielectric function in the presence of a static field 
${\cal E}_0$ as
follows. The system is prepared at $t=0$ in the stationary state polarized
by a field ${\cal E}_0+\Delta{\cal E}$, with $|\Delta{\cal E}|<<|{\cal E}_0|$. 
By using a field of magnitude below the
critical field, we are able to find that state by 
minimizing the energy. 
For $t>0$ we let the system evolve in time in the presence of the target field 
${\cal E}_0$.
Let $P_{\rm static}[{\cal E}_0]$ be static polarization of the system under the
field ${\cal E}_0$. The polarization response to the step-function
discontinuity in ${\cal E}(t)={\cal E}_0+\Delta {\cal E}\,\theta(-t)$ is
$\Delta P(t)=P(t)-P_{\rm static}[{\cal E}_0]$. 
To obtain the frequency-dependent response we need the Fourier
transform of $\Delta P(t)$ for $t>0$ only:
\begin{equation}
\Delta P(\omega)=\int_0^{+\infty}\,\Delta P(t)\,
e^{(i\omega-\delta)t}\,dt,
\end{equation}
where a damping factor $\delta$ has been introduced as an approximate way to
account for level broadening.\cite{tsolakidis02}
To linear order in $\Delta {\cal E}$ the susceptibility is
\begin{equation}
{\rm Re}\,\chi^{[{\cal E}_0]}(\omega)=
\left. \frac{dP_{\rm static}[{\cal E}]}{d{\cal E}}\right|_{{\cal E}={\cal E}_0} -
\frac{\omega}{\Delta \cal E}\,{\rm Im}\,\Delta P(\omega),
\end{equation}
\begin{equation}
{\rm Im}\,\chi^{[{\cal E}_0]}(\omega)=\frac{\omega}{\Delta \cal E}\,
{\rm Re}\,\Delta P(\omega).
\end{equation}

With this real-time approach the need to perform a summation over 
conduction-band states is circumvented.
Previous real-time, scalar potential
approaches\cite{tsolakidis02,yabana99} were restricted to finite systems,
since it was unclear how to evaluate the dynamic macroscopic polarization of an
extended system. A real-time, vector-potential
scheme valid for bulk systems was proposed in Ref.~\onlinecite{bertsch00}.

We validate our method by comparing in Fig.~\ref{fig6}
the ground-state susceptibility 
with the analytic Kubo-formula (sum-over-states)
result, using in both cases the same 
broadening $\delta$ and $k$-point mesh. Also shown in Fig.~\ref{fig6} is the
susceptibility in the presence of a ${\cal E}_0=0.05$ bias field,
displaying the Franz-Keldysh effect: an
absorption tail below the gap caused by photon-assisted tunneling, and 
oscillations above the gap.\cite{yu96} 
The Franz-Keldysh oscillations become more widely spaced with increasing
${\cal E}_0$. This is illustrated in the inset of 
Fig.~\ref{fig6}, where we compare them for ${\cal E}_0=0.05$ and
${\cal E}_0=0.03$.


\section{Summary}

The work of King-Smith and Vanderbilt demonstrated that the bulk electronic
polarization, defined in terms of the current flowing during the
{\it adiabatic} evolution of an insulating system in a {\it vanishing
macroscopic electric field}, could be related to a Berry's phase
defined over the manifold of occupied Bloch states.\cite{ksv93}
We have generalized this result by considering the time evolution of
an initially insulating electron system under the very general
Hamiltonian (\ref{eq:hamiltonian}), where
the lattice-periodic part $\hat{H}^0(t)$ and the homogeneous
electric field $\ee(t)$ may have an arbitrarily strong and
rapid variation in time. In the absence of scattering, we have proved that the
integrated current $\Delta {\bf P}=\int {\bf J}(t)\,dt$ is still
given by the King-Smith--Vanderbilt formula, but written in terms
of the instantaneous Bloch-like solutions of the time-dependent
Schr\"odinger equation.  The coherent dynamic polarization ${\bf P}(t)$ was
interpreted as a nonadiabatic geometric phase.\cite{aa87} These
generalizations of the theory allowed us to justify recent developments in 
which 
the energy functional $E$ of Nunes and Gonze\cite{nunes01} has been used as the
basis for direct DFT calculations of insulators in a static
homogeneous electric field.\cite{souza02,umari02} 
The limitation of those methods to fields of magnitude smaller than a 
$\Delta \kk$-dependent
critical field that vanishes in the thermodynamic limit 
has been removed: we have shown numerically that quasistationary states in 
finite fields exist for arbitrarily dense $k$-point meshes, and can be obtained
by solving the time-dependent Shr\"odinger equation for a slowly-increasing
field.
The present method also provides a
convenient framework for the computation of coherent time-dependent
excitations in insulators.   
As an example, the dielectric function
was calculated for a tight-binding model by considering the response
to a step-function discontinuity in $\ee(t)$, illustrating
effects such as photon-assisted tunneling and Franz-Keldysh oscillations.
A full {\it ab initio} implementation within the framework of time-dependent
density-functional theory should be possible.

\begin{acknowledgements}
This work was supported by NSF Grant DMR-0233925 and by DARPA/ONR Grant
N00014-01-1-1061.  We wish to thank M.~H.~Cohen for many useful discussions.
\end{acknowledgements}


\appendix

\section{Crystal-momentum representation}
\label{app:pol}

The 
introduction of linear scalar potentials in crystals is usually discussed in
the language of the crystal-momentum representation (CMR).\cite{blount62}
Instead, we have used the Berry-phase theory of polarization, and the purpose 
of this Appendix is to show how to switch from one to the other. The
CMR uses 
as a basis the eigenstates $\ket{\psi_{\kk m}}$ of $\hat{H}^0$ with eigenvalues
$E_{\kk m}$. In accordance with Eq.~(\ref{eq:norm_conv}) we assume that 
$|\psi_{\kk m}(\rr)|^2$
integrates to one over the unit cell volume $v$. That 
implies\cite{jones73,foot:normalization}
\begin{equation}
\label{eq:normalization}
\bra{\psi_{\kk m}}\psi_{\kk' l}\rangle\equiv\int \psi_{\kk m}^*(\rr)
\psi_{\kk' l}(\rr)\,d\rr=\Omega_B\delta(\kk-\kk')\delta_{ml}.
\end{equation}
The CMR expansion of the identity operator is
\begin{equation}
\hat{\bf 1}=\Omega_B^{-1}\sum_{m=1}^\infty\int d\kk\,
\ket{\psi_{\kk m}}\bra{\psi_{\kk m}},
\end{equation}
so that a general 
one-electron state $\ket{\phi}$ is expanded as
\begin{equation}
\label{eq:cmr}
\ket{\phi}=\hat{\bf 1}\ket{\phi}=
\sum_{m=1}^\infty\,\int\,d\kk' f_{\kk' m}\ket{\psi_{\kk' m}},
\end{equation}
where $f_{\kk' m}=\Omega_B^{-1} \bra{\psi_{\kk' m}}\phi\rangle$.
For the occupied Bloch-like states $\ket{\phi_{\kk n}}$ in a WR manifold, the 
CMR wave function $f_{\kk' m}(\kk,n)$ takes the form
\begin{equation}
f_{\kk' m}(\kk,n)=c_{\kk',nm}\delta(\kk'-\kk) 
\end{equation}
with $\sum_{m=1}^\infty |c_{\kk,nm}|^2=1$,
which leads to Eq.~(\ref{eq:linear_comb}).

\subsection{Current and the CMR velocity operator}

The velocity operator (\ref{eq:velocity_b})
is diagonal in $\kk$ and is conveniently split into a sum of two 
operators, one diagonal and the other off-diagonal in the band
index:\cite{blount62}
\begin{equation}
\label{eq:velocity_cmr}
\hat{\bf v}=\hat{\bf v}^{\rm d}+\hat{\bf v}^{\rm od}.
\end{equation}
The matrix elements of $\hat{\bf v}^{\rm d}$ are
\begin{equation}
\langle \psi_{\kk m}|\hat{\bf v}^{\rm d}|\psi_{\kk' l}\rangle=
\Omega_B\delta(\kk-\kk')\delta_{ml}{\bf v}^{\rm d}_{\kk m},
\end{equation} 
where
\begin{equation}
\label{eq:vel_intra}
{\bf v}^{\rm d}_{\kk m}
= \frac{1}{\hbar}\,\partial_\kk E_{\kk m}.
\end{equation}
The matrix elements of $\hat{\bf v}^{\rm od}$ are
\begin{equation}
\langle \psi_{\kk m}|\hat{\bf v}^{\rm od}|\psi_{\kk' l}\rangle=
\Omega_B\delta(\kk-\kk'){\bf v}^{\rm od}_{\kk,ml},
\end{equation}
where
\begin{equation} 
\label{eq:vel_inter}
{\bf v}^{\rm od}_{\kk,ml}
=\frac{i}{\hbar}\,{\bf X}_{\kk,ml}\,[E_{\kk m}-E_{\kk l}]
\end{equation}
and we have defined the hermitian matrix
\begin{equation}
\label{eq:x_matrix}
X_{\kk,ml}^\alpha=i\,\langle 
                      u_{\kk m}|\partial_{k_\alpha} u_{\kk l}
                    \rangle,
\end{equation}
which is analogous to Eq.~(\ref{eq:connection}) for the $\ket{v_{\kk n}}$.

The current, Eq.~(\ref{eq:current}), is split into intraband and interband 
parts,
\begin{equation}
\label{eq:total_current}
\jj(t)=\jj_{\rm intra}(t)+\jj_{\rm inter}(t).
\end{equation}
Writing the density matrix as
\begin{equation}
\label{eq:cmr_dm}
\langle \psi_{\kk m}|\hat{n}|\psi_{\kk' l}\rangle=
\Omega_B\delta(\kk-\kk')n_{\kk,ml},
\end{equation}
where
\begin{equation}
\label{eq:dm_cmr}
n_{\kk,ml}= 
\sum_{n=1}^M\,c_{\kk,nm}\,{[c_{\kk,nl}]}^{*},
\end{equation}
we find
\begin{equation}
\label{eq:intra_current}
\jj_{\rm intra}= -\frac{e}{v}\,{\rm Tr}_c\big(\hat{n}\hat{\bf v}^{\rm d}\big)
=\frac{-e}{(2\pi)^3} \, \sum_{m=1}^{\infty}\,\int\,d\kk\,
n_{\kk,mm} \, {\bf v}^{\rm d}_{\kk m}
\end{equation}
and
\begin{equation}
\label{eq:inter_current}
\jj_{\rm inter}=-\frac{e}{v}\,{\rm Tr}_c\big(\hat{n}\hat{\bf v}^{\rm od}\big)
=\frac{-e}{(2\pi)^3} \sum_{m,l=1}^{\infty} \, \int \, d\kk\,
n_{\kk,ml} {\bf v}^{\rm od}_{\kk,lm}.
\end{equation} 
In the above we used the CMR form of Eq.~(\ref{eq:trace}),
\begin{equation}
\label{eq:trace_cmr}
{\rm Tr}_c(\hat{\cal O})=\Omega_B^{-1}\sum_{m=1}^\infty\,
\int d\kk\,\frac{1}{N}\bra{\psi_{\kk m}}\hat{\cal O}\ket{\psi_{\kk m}},
\end{equation}
where $N$ should be taken to signify 
$\Omega_B\delta(0)$.\cite{foot:normalization}

Plugging (\ref{eq:linear_comb}) into (\ref{eq:curr}) yields, after some
manipulations,
Eqs.~(\ref{eq:total_current}), (\ref{eq:intra_current}), and
(\ref{eq:inter_current}), confirming that
the Berry-phase polarization correctly accounts for both 
intraband and interband contributions.
It is instructive to consider some particular cases. The adiabatic current
$\jj=(d\pp/d\lambda)\dot{\lambda}$
discussed in Refs.~\onlinecite{ksv93,thouless83} is purely interband.
If the perturbation is a sinusoidal electric field, the linear 
response is again a purely interband current, while the nonlinear response 
has also an intraband component.\cite{sipe00,lambrecht00}

\subsection{Polarization and the CMR position operator}

Along the same lines, one can show that the Berry-phase expression for
$\pp$ is consistent with the CMR position operator, which takes the 
form\cite{blount62}
\begin{equation}
\label{eq:cmr_x}
\bra{\psi_{\kk m}}\hat{\rr}\ket{\psi_{\kk'l}}=-i\Omega_B\partial_{\kk'}
\delta(\kk'-\kk)\delta_{mn}+\Omega_B\delta(\kk'-\kk){\bf X}_{\kk,ml}.
\end{equation}
Combined with Eqs.~(\ref{eq:cmr_dm}) and (\ref{eq:trace_cmr}) this yields
\begin{equation}
\pp=-\frac{e}{v}{\rm Tr}_c(\hat{n}\hat{\rr})
=\frac{-e}{(2\pi)^3}\sum_{m,l=1}^\infty\int d\kk\,
n_{\kk,ml}{\bf X}_{\kk,lm},
\end{equation}
which is the same results one gets from inserting the CMR expansion
(\ref{eq:linear_comb}) into the nonadiabatic Berry-phase formula 
(\ref{eq:dyn_pol}). The linear character of $\hat{\rr}$ is reflected in the 
above equation being defined only up to a quantum of polarization.

\subsection{CMR dynamical equations}

In the case where $\hat{H}^0$ (and hence the CMR basis) is constant in time,
plugging (\ref{eq:linear_comb}) into the TDSE (\ref{eq:tdse_continuum}) yields
the CMR form of the Schr\"odinger equation,\cite{adams57,callaway91}
\begin{equation}
\label{eq:tdse_cmr}
i\hbar\,\dot{c}_{\kk m}=
\big(
  E_{\kk m}+ie\ee\cdot{\cal D}_\kk
\big) c_{\kk m},
\end{equation}
where we have simplified $c_{\kk,nm}$ to $c_{\kk m}$ and defined
\begin{equation}
{\cal D}_\kk c_{\kk m}=\partial_\kk c_{\kk m}-
i\sum_{l=1}^\infty\,{\bf X}_{\kk l}c_{\kk l},
\end{equation}
which is reminiscent of the covariant derivative, Eq.~(\ref{eq:cov_der})
(but note the difference in the sign of the last term).
It is customary to write Eq.~(\ref{eq:tdse_cmr}) as
\begin{equation}
\label{eq:tdse_cmr_b}
i\hbar\,\dot{c}_{\kk m}=
\big(
  E_{\kk m}^{(1)}+ie\ee\cdot\partial_\kk
\big) c_{\kk m}+e\ee\,\cdot\,\sum_{l\not=m}^\infty\,c_{\kk l}
{\bf X}_{\kk,ml},
\end{equation}
where
\begin{equation}
\label{eq:shifted_band}
E_{\kk m}^{(1)}=E_{\kk m}+e\ee\cdot {\bf X}_{\kk,mm}
\end{equation}
is a shifted energy eigenvalue. $E_{\kk m}^{(1)}$
is identical to Eq.~(\ref{eq:energy_k_density}) except that
$\ket{v_{\kk m}}$ has been replaced by the zero-field eigenstate 
$\ket{u_{\kk m}}$.
Upon averaging over $\kk$ the last term on the right-hand-side becomes
the first-order shift in total energy,
$-v\pp_0\cdot\ee$, where $\pp_0$ is the spontaneous Berry-phase polarization.

In general the above TDSE has no stationary solutions.
Approximate solutions --~the Wannier-Stark states~-- result from 
restricting the wavepacket dynamics to a single band
(the semiclassical approximation). That is achieved by
dropping the sum on the right-hand-side of Eq.~(\ref{eq:tdse_cmr_b}),
which is responsible for interband tunneling.\cite{callaway91,kane59}

Finally, combining (\ref{eq:dm_cmr}) and (\ref{eq:tdse_cmr}) produces the
dynamical equation for the CMR density matrix:
\begin{eqnarray}
\label{eq:sbe}
i\hbar\,\dot{n}_{\kk,nm}&=&
(E_{\kk n}-E_{\kk m})n_{\kk,nm}+ie\ee\cdot\partial_\kk n_{\kk,nm} \nonumber \\
&-&e\ee\,\cdot\,\sum_{l=1}^\infty\,
\big(
  n_{\kk,nl}{\bf X}_{\kk,lm}-{\bf X}_{\kk,nl}n_{\kk,lm}
\big).\nonumber \\
\end{eqnarray}
A closely related form has been used to study the nonlinear optical 
susceptibilities of semiconductors.\cite{aversa95,lambrecht00}


\section{Covariant derivative and related operators}
\label{app:cov}

In Sec.~\ref{sec:td_continuum} we introduced a modified TDSE that contains
the multiband covariant derivative $\widetilde{\partial}_\kk$, 
Eq.~(\ref{eq:cov_der}), that was instrumental for making contact with
the discrete-$k$ dynamical equations of Sec.~\ref{sec:td_discrete}.
Here we summarize the properties of the covariant derivative and other closely
related operators.

The covariant derivative $\widetilde{\partial}_\kk\ket{v_{\kk n}}$ 
of an occupied state transforms
in the same way as that state under a gauge transformation, 
Eq.~(\ref{eq:gauge_transf}):
\begin{equation}
\label{eq:gauge_cov_der}
\widetilde{\partial}_\kk\ket{v_{\kk n}}\rightarrow
\sum_{m=1}^M\,U_{\kk,mn}\,\widetilde{\partial}_\kk\ket{v_{\kk m}}.
\end{equation}
Moreover, it is orthogonal to the occupied subspace at $\kk$,
\begin{equation}
\label{eq:orthogonal}
\bra{v_{\kk m}}\widetilde{\partial}_\kk v_{\kk n}\rangle=0.
\end{equation}
Recalling that parallel transport is characterized by
$\bra{v_{\kk n}} \partial_\kk v_{\kk n}\rangle=0$, for
$m=n$ this relation shows that $\widetilde{\partial}_\kk$
acting in an arbitrary gauge gives the same result as $\partial_\kk$
acting in the parallel-transport gauge that shares the same states
at $\kk$. In the discretized form (\ref{eq:cov_discrete}) the property 
(\ref{eq:gauge_cov_der}) is a consequence of Eq.~(\ref{eq:gauge_cov}), and the
property (\ref{eq:orthogonal}) is a consequence of Eq.~(\ref{eq:duality}).
Like $i\partial_\kk$, $i\widetilde{\partial}_\kk$ is hermitian.
By this we mean that
its matrix representation in an orthonormal basis 
(e.g., the $\ket{v_{\kk n}}$, $n=1,\ldots,M$ complemented by a set of 
unoccupied states $\ket{c_{\kk j}}$) is hermitian. This is closely related
to the hermiticity of the matrix $A_{\kk}^\alpha$ defined in
Eq.~(\ref{eq:connection}). Finally, note that
\begin{equation}
\label{eq:related_ops}
i\widetilde{\partial}_\kk\ket{v_{\kk n}}=
i\hat{Q}_\kk\partial_\kk\ket{v_{\kk n}}
=i\hat{Q}_\kk\partial_\kk\hat{P}_\kk\ket{v_{\kk n}},
\end{equation}
i.e., the action of $i\widetilde{\partial}_\kk$ on an occupied state is 
identical to that of $i\hat{Q}_\kk\partial_\kk$ and
$i\hat{Q}_\kk\partial_\kk\hat{P}_\kk$. They differ in how they
act on the unoccupied states. Unlike
$i\widetilde{\partial}_\kk$, the other two
are not hermitian: for instance, 
$
(i\hat{Q}_\kk\partial_\kk\hat{P}_\kk)^{\dagger}\ket{v_{\kk n}}=0$.
It follows from these considerations that 
Eq.~(\ref{eq:tdse_continuum_cov}) can be recast as
\begin{equation}
\label{eq:tdse_continuum_cov_mod}
i\hbar\ket{\dot{v}_{\kk n}}=\big[\hat{H}^0_\kk+
e\ee\cdot\big(i\hat{Q}_\kk\partial_\kk\hat{P}_\kk+
\text{h.c.}
\big)
\big]\ket{{v}_{\kk n}}.
\end{equation}
This is the form of the TDSE to which 
Eqs.~(\ref{eq:td_discrete_b})-(\ref{eq:t_k}) reduce in the continuum-$k$
limit, since 
\begin{equation}
\hat{\rm w}_\kk\simeq ie\ee\cdot\hat{Q}_\kk\partial_\kk\hat{P}_\kk
\end{equation}
(compare with Eq.~(\ref{eq:cov_dis})).


\section{Gradient of the energy functional}

The purpose of this Appendix is to obtain expressions for the
derivatives of the two terms in the energy functional
of Eq.~(\ref{eq:energy}) with respect to the occupied Bloch-like
states in the discrete-$k$ case.  The results have been used
in Secs.~\ref{sec:td_discrete} and \ref{sec:stat_formulation} for
the discussion of the time-dependent evolution equations and the
stationary solutions, respectively.

\subsection{Band-structure contribution}
\label{app:grad-a}

To find the gradient $\delta E/\bra{\delta v_{\kk n}}$ of the energy
functional (\ref{eq:energy}), let us isolate the terms that depend on 
$\bra{v_{\kk n}}$. Using (\ref{eq:projector}) the zero-field part 
(\ref{eq:energy0_discrete}) can be expressed as
$E^0=(1/N)\,\sum_\kk\,{\rm tr}[\hat{P}_\kk\hat{H}^0_\kk]$, so that
\begin{equation}
\frac{\delta E^0}{\bra{\delta v_{\kk n}}}=\frac{1}{N}
\frac{\delta {\rm tr}[\hat{P}_\kk\hat{H}^0_\kk]}{\bra{\delta v_{\kk n}}}.
\end{equation}
In order to allow for arbitrary variations of $\bra{v_{\kk n}}$,
even those for which the $\bra{v_{\kk n}}$ do not remain orthonormal,
we write
\begin{equation}
\label{eq:proj_nonorthonormal}
\hat{P_\kk}=\sum_{m,n=1}^M {({\cal S}_\kk^{-1})}_{mn} 
\ket{v_{\kk m}} \bra{v_{\kk n}},
\end{equation}
where ${\cal S}_{\kk,mn}=\langle v_{\kk m}|v_{\kk n}\rangle$. Dropping the 
subscript $\kk$,
\begin{eqnarray}
&{\delta{\rm tr}[\hat{P}\hat{H}^0]}&=
{{\rm tr}[(\delta\hat{P})\hat{H}^0]} \nonumber \\
&&=\sum_{m,n}\,({\cal S}^{-1})_{mn}
\Big[\langle v_n|\hat{H}^0|\delta v_m\rangle+
    \langle \delta v_n|\hat{H}^0|v_m\rangle
\Big]
\nonumber \\
&&\qquad+\sum_{m,n}\,\langle v_n|\hat{H}^0|v_m\rangle\,
\delta({\cal S}^{-1})_{mn}.
\end{eqnarray}
Using $\delta({\cal S}^{-1})=-{\cal S}^{-2}\,\delta{\cal S}$ and
$
\delta {\cal S}_{mn}=
\langle v_m|\delta v_n\rangle+\langle \delta v_m|v_n\rangle,
$
and evaluating at ${\cal S}={\bf 1}$, we arrive at
\begin{equation}
\label{eq:grad_zerofld}
\frac{\delta {\rm tr}[\hat{P}_\kk\hat{H}^0_\kk]}{\bra{\delta v_{\kk n}}}=
\hat{Q}_\kk\,\hat{H}_\kk^0\,\ket{v_{\kk n}}.
\end{equation}
Thus the consequence of expressing $\hat{P}_\kk$ as 
(\ref{eq:proj_nonorthonormal}) 
instead of (\ref{eq:projector}) is to 
render the gradient orthogonal to the occupied manifold at $\kk$.
(When we derived the dynamical equation (\ref{eq:td_discrete_a})
using (\ref{eq:euler-lagrange_b}), the
gradient of $E^0$ was not orthogonalized, which is why the
dynamics did not follow parallel transport (see Sec.~\ref{sec:integration})).

\subsection{Polarization contribution}
\label{app:grad-b}

To find the gradient of the field-coupling term $-v\ee\cdot\pp$
we need $\delta\overline{\Gamma}_i/\bra{\delta v_{\kk n}}$. Let us start by
recasting Eq.~(\ref{eq:discrete_phase}) as
\begin{equation}
\overline{\Gamma}_i=
\frac{1}{N^\perp_i}\,\sum_{l=1}^{N^\perp_i}\,\sum_{j=0}^{N_i^\parallel-1}\,
\phi(\kk_j^{(i)},\kk_j^{(i)}+\Delta \kk_i),
\end{equation}
where we have defined the phase
\begin{equation}
\label{eq:phase}
\phi(\kk,\kk')=-{\rm Im}\,{\rm ln}\,{\rm det}\,
S(\kk,\kk').
\end{equation}
Using $\phi(\kk',\kk)=-\phi(\kk,\kk')$, this becomes
\begin{equation}
\label{eq:gamma_avg}
\overline{\Gamma}_i=\frac{1}{N^\perp_i}\,\sum_{\sigma=\pm 1}\,\sigma\,
\phi(\kk,\kk i\sigma)+\ldots,
\end{equation}
where only the terms depending on $\bra{v_{\kk n}}$ were written explicitly.
Hence
\begin{equation}
\frac{\delta \overline{\Gamma}_i}{\bra{\delta v_{\kk n}}}=
\frac{1}{N^\perp_i}\,\sum_{\sigma=\pm 1}\,\sigma\,
\frac{\delta}{\bra{\delta v_{\kk n}}}\phi(\kk,\kk i\sigma).
\end{equation}
The phase $\phi(\kk,\kk')$ can be expressed as
\begin{eqnarray}
\label{eq:phase_b}
\phi(\kk,\kk')&=&-{\rm Im}\,{\rm tr}\,{\rm ln}\,
S(\kk,\kk')\nonumber\\
&=&\frac{i}{2}\,{\rm tr}\,{\rm ln}\,S(\kk,\kk')\,-\,
\frac{i}{2}\,{\rm tr}\,{\rm ln}\,S(\kk',\kk).
\end{eqnarray}
For an arbitrary non-singular matrix $A$ we have
\begin{eqnarray}
\delta {\rm tr}\,{\rm ln}A&=&{\rm tr}\,{\rm ln}(A+\delta A)-{\rm tr}\,
{\rm ln}(A)\nonumber \\
&=&{\rm tr}\,{\rm ln}[(A+\delta A)A^{-1}]=
{\rm tr}\,{\rm ln}[{\bf 1}+(\delta A)A^{-1}]\nonumber\\
&=&{\rm tr}[A^{-1}\,\delta A]+{\cal O}(\delta A^{2}),
\end{eqnarray}
so that
\begin{eqnarray}
\frac{\delta {\rm tr}\,{\rm ln}S(\kk,\kk')}{\bra{\delta v_{\kk n}}}&=&
{\rm tr}
\left[
  S^{-1}(\kk,\kk')\,\frac{\delta S(\kk,\kk')}{\bra{\delta v_{\kk n}}}
\right]\nonumber \\
&=&\sum_{m=1}^M\,S^{-1}_{mn}(\kk,\kk')\,\ket{v_{\kk' m}}\nonumber\\
&=&\ket{\widetilde{v}_{\kk'n}}.
\end{eqnarray}
The corresponding derivative of the last term of (\ref{eq:phase_b})
vanishes since $S(\kk',\kk)=\langle v_{\kk'n}\vert v_{\kk n}\rangle$
does not contain $\bra{v_{\kk n}}$ as a bra.  We thus arrive at
\begin{equation}
\label{eq:grad_phase}
\frac{\delta \phi(\kk,\kk')}{\bra{\delta v_{\kk n}}}=\frac{i}{2}\,
\ket{\widetilde{v}_{\kk'n}},
\end{equation}
which combined with (\ref{eq:gamma_avg}) gives 
\begin{equation}
\label{eq:grad_berry_phase}
\frac{\delta \overline{\Gamma}_i}{\bra{\delta v_{\kk n}}}=
\frac{i}{2N^\perp_i}\sum_{\sigma=\pm1}\,\sigma |\widetilde{v}_{\kk i\sigma,n}
\rangle.
\end{equation}
This is automatically orthogonal to the occupied manifold at $\kk$.
(See Refs.~\onlinecite{nunes01,sai02} for alternative derivations.)
Collecting terms and using Eq.~(\ref{eq:ket_w_k}),
we obtain Eq.~(\ref{eq:gradient_energy}) for the gradient of the full
energy functional $E$.


\section{Discretized formula for the current}
\label{app:discr}

Just as the macroscopic polarization $\pp$
is evaluated in practice via a finite-difference formula on a mesh of $k$ 
points, the same can be done for the macroscopic current $\jj=d\pp/dt$.
The invariance of Eq.~(\ref{eq:curr}) under the
replacement $\partial_\kk\rightarrow\widetilde{\partial}_\kk$ allows us to then
use the discretization rule (\ref{eq:cov_discrete}), 
leading to
\begin{equation}
\label{eq:curr_discrete}
\jj=\frac{e}{4\pi\hbar v}\,\sum_{n,\kk}\,\sum_{i=1}^3\,\sum_{\sigma=\pm 1}\,
\frac{\sigma}{N_i^\perp}\,
\langle v_{\kk n}|\hat{H}_\kk^0|\widetilde{v}_{\kk i\sigma,n}\rangle\aa_i
+{\rm c.c.},
\end{equation}
which is significantly easier and cheaper to compute than the spatial average 
of the microscopic current.
We have checked numerically on our one-dimensional tight-binding model
that Eq.~(\ref{eq:curr_discrete}) yields, for small $\Delta t$, the same result
as $[P(t+\Delta t)-P(t)]/\Delta t$ computed with the
discretized Berry-phase formula. 

The same strategy as outlined above can be used to derive a discretized
formula for the Berry curvature (\ref{eq:curvature}) summed over
bands, which is also invariant under 
$\partial_\kk\rightarrow\widetilde{\partial}_\kk$. 
This may be useful in other contexts, such as semiclassical wavepacket dynamics
in crystals.\cite{chang96}



\begin{thebibliography}{0}
\bibitem{foot:mistake} An incorrect interpretation was given
in Ref.~\onlinecite{nunes01}
to the term $ie{\ee}\cdot\partial_\kk\ket{{v}_{\kk n}}$. It does not come
from the ``periodic part'' of the position operator, but rather from
the full position operator, as can be seen from the present derivation.
\bibitem{foot:wavepacket} After removing the ``band'' indices
Eq.~(\ref{eq:orthonormality}) becomes the well-known equation
for the $k$-space dynamics of an electronic wavepacket 
$|\psi\rangle=\int |\phi_{\kk}\rangle\,d\kk$:
$\dot{n}_{\kk}=(e/\hbar)\ee\cdot\partial_{\kk}n_{\kk}$, where
$n_{\kk}=\langle v_{\kk}|v_{\kk} \rangle$. Eq.~(\ref{eq:tdse_continuum})
is also valid for such wavepackets, even if the derivation given in the text
was for WR manifolds.
\bibitem{foot:frustrated_metal}  Such a coherent ``frustrated metal'' state
can exist in the early moments following a strong femtosecond laser-pulse
excitation of a semiconductor, but it will eventually be destroyed by thermal 
scattering.\cite{mazur02}
\bibitem{explan-notanom} 
The resemblance of Eqs.~(\ref{eq:vanishing_part})-(\ref{eq:curvature}) to the
so-called anomalous current\cite{chang96} in semiclassical wavepacket
dynamics is misleading, since the Berry
curvature entering the anomalous current pertains to the band states
$\ket{u_{\kk n}}$, not to the $\ket{v_{\kk n}}$. 
In fact, it can be shown that
the anomalous current is contained in the other term,
Eq.~(\ref{eq:curr}), once it is properly modified for wavepackets.
\bibitem{foot:thouless} It may seem surprising that even though we use a 
discrete mesh of
$k$-points --~ which effectively turns the system into a finite torus~-- 
our calculations display exact quantization of adiabatic particle transport. 
According to Thouless,\cite{thouless83} that  should only
occur in the thermodynamic (continuum-$k$) limit. This has to do with our
using the discretized Berry-phase formula (\ref{eq:discrete_phase}),
which is not fully consistent with the velocity operator (\ref{eq:velocity})
on the finite torus (we recall that the derivation in 
Sec.~\ref{sec:dyn_pol_der} was done in the
continuum-$k$ limit). From this viewpoint it seems preferable to regard the 
discretization procedure as a numerical approximation
to the continuum-$k$ limit, 
rather than as an exact calculation on the finite torus.
\bibitem{foot:insulating-like} In the limit where the number $N$ of $k$ points 
is large,
$\xi$ is finite for the ground state of insulators and diverges for that of
metals.\cite{souza00} 
For finite $N$ it cannot really diverge. In particular,
when $\xi=(\Omega_{\rm I}/M)^{1/2}$ is evaluated using Eq.~(34)
of Marzari and Vanderbilt\cite{mv97} it cannot exceed, in one dimension, 
$aN/2\pi$, well above the values that we observe
in all our simulations.
\bibitem{foot:normalization} Since $\delta(\kk-\kk')$ has units of real-space
volume, the right-hand side of Eq.~(\ref{eq:normalization}) is dimensionless.
In the discrete-$k$ case the integral on the left-hand side is over the
supercell volume $V=Nv$, and $\Omega_B\delta(\kk-\kk')$ is replaced 
by $N\delta_{\kk \kk'}$.
\end{thebibliography}
\end{document}